\documentclass[usenatbib]{mnras}
\usepackage[T1]{fontenc}
\usepackage{ae,aecompl}

\usepackage{graphicx}	
\usepackage{amsmath}	
\usepackage{amssymb}	
\usepackage{array}

\usepackage{multirow}
\usepackage{multicol}
\usepackage{blindtext}
\newcolumntype{?}{!{\vrule width 1pt}}
\newcommand{\Msun}{\,{\rm M}$_{\odot}$\,}
\newcommand{\Mpch}{\,{\rm Mpc}\,\ifmmode h^{-1}\else $h^{-1}$\fi}
\newcommand{\kpch}{\,{\rm kpc}\,\ifmmode h^{-1}\else $h^{-1}$\fi}

\title[Dark matter halo shapes in Auriga] {Dark matter halo shapes in the Auriga simulations} 
\author[Prada et al.]{
\parbox[t]{\textwidth}{
{Jesus Prada,$^{1}$\thanks{E-mail: jd.prada1760@uniandes.edu.co}}
{Jaime E. Forero-Romero,$^{1}$\thanks{E-mail: je.forero@uniandes.edu.co}}
{Robert J. J. Grand,$^{2}$}
{R\"udiger Pakmor,$^{2}$}
{Volker Springel$^{2}$}
}
\\\\
$^{1}$Departamento de F\'isica, Universidad de los Andes, Cra. 1 No.
18A-10, Edificio Ip, Bogot\'a, Colombia\\
$^{2}$Max-Planck-Institut f\"ur Astrophysik, Karl-Schwarzschild-Str. 1, D-85741 Garching, Germany\\
}

\date{Accepted XXX. Received YYY; in original form ZZZ}

\pubyear{2019}

\begin{document}
\label{firstpage}
\pagerange{\pageref{firstpage}--\pageref{lastpage}}
\maketitle

\begin{abstract}
We present shape measurements of Milky Way-sized dark matter halos at
redshift $z=0$ in a suite of 30 zoom simulations from the Auriga
project. 
We compare the results in full magnetohydrodynamics against dark
matter only simulations and find a strong influence of baryons 
in making dark matter haloes rounder at all radii compared to their dark 
matter only counterparts.
At distances $\lesssim 30$ kpc, rounder dark matter distributions
correlate with extended massive stellar discs and low core gas 
densities.  
We measure the alignment between the halo and the disc shapes at
different radii and find a high degree of alignment at all radii for most 
of the galaxies.
In some cases the alignment significantly changes as a function of
radius implying that the halo shape twists; 
this effect correlates with recently formed bulges and is almost
absent in the dark matter only simulations.  
In a comparison against observational constraints we find that $20\%$
of halos in our sample are consistent with observational results derived
from the Pal 5 stream that favours an almost spherical shape.
Including baryons is a required element to achieve this
level of agreement. In contrast, none of the simulations (neither dark matter
only nor with baryons) match the constraints derived from the
Sagittarius stream that favour an oblate dark matter halo.
\end{abstract}

\begin{keywords}
galaxies: evolution --- galaxies: formation --- galaxies: haloes ---
dark matter
\end{keywords}



\section{Introduction}

Our physical picture of the Universe as a whole has been shaped by
accurate observations and modeling of our own Galaxy. 
For instance, the study of the Milky Way's (MW) morphology and its
separation into different kinematic components, such as 
disc and bulge, has been used to support the existence of a Dark Matter (DM) component surrounding the Galaxy
\citep{2000MNRAS.311..361O,2009PASJ...61..227S,2010JCAP...08..004C,2013ApJ...779..115B,Iocco15}. 
Explaining the matter budget and kinematic state of the MW also puts constraints on the wider cosmological context, besides testing specific models of galactic evolution.

A detailed study of the full three-dimensional MW
gravitational potential could constrain the properties of the DM halo surrounding our Galaxy and even help to pin down the nature of the DM particle. For instance, the study of fossil records of stellar streams resulting from infalling globular clusters or
satellite galaxies that got tidally disrupted by the gravitational
potential of the MW could be translated into tight constraints 
both on the shape of the dark matter halo in the outer regions of our Galaxy \citep{1998ApJ...495..297J,1999MNRAS.307..495H, 1999MNRAS.307..877T} and the properties of dark matter substructures, which also depend on the microphysics of a given DM particle candidate.

In any case, the observational constraints on the gravitational potential shape must be confronted against the expectations from different galaxy formation models in the full cosmological context.
For instance, in the current paradigm of a Cold Dark Matter
(CDM) dominated universe, galaxies are expected to be hosted by
triaxial DM halos. To what extent are these CDM expectations born out in observations of our Galaxy? 

This question has been challenging to address for many years because it has been difficult to produce realistic galactic discs in simulations within the CDM context \citep{1997ApJ...478...13N}.
It is now understood that a prerequisite for successful disc formation is an adequate 
accounting of strong baryonic feedback effects, such as stellar feedback and black hole feedback, because they play an important role in forming late-time galaxies resembling the MW.
Along the way, different numerical experiments have also shown that the baryonic
effects also impact the DM halo shape, making it rounder
than it would otherwise be in a DM-only simulation
\citep{Dubinski94,Debattista08,Kazantzidis10,Abadi10,Bryan13,Chua19,Artale19}. 
Another important aspect of the disc-halo relationship, namely the radial evolution of the
alignment between the stellar disc and the DM halo, has also been explored in some
simulations \citep{Bailin05,Debattista13}.

Observational constraints on the dark matter halo shape
have partly been in mutual tension, and have been rapidly evolving during the last decade, hinting that they are affected by substantial systematic uncertainties.
One can find studies favouring prolate
\citep{Banerjee_and_Chanda_2011,Bowden_et_al._2016},  oblate
\citep{LM10,Deg_and_Widrow_2013,Vera-Ciro_and_Helmi_2013} and
spherical configurations \citep{Bovy16}.  
Constraints from modeling stellar streams discard the prolate
configuration \citep{LM10,Pearson_et_al._2015,Bovy16} although some other studies
still question whether stellar streams can be used to constrain the halo
shape once certain assumptions, such as the density profile, are relaxed
\citep{Ibata_et_al._2013}.

In this paper, we analyze the internal DM halo shape in thirty MW-type
galaxies  from the Auriga project, which are state-of-the-art hydrodynamical simulations of galaxy formation.
They have large enough numerical resolution, include an explicit
cosmological context, and have appropriate feedback physics to produce
realistic MW discs \citep{auriga,GBG18,GHF18}.
Furthermore, the relatively large number of simulated systems allows us
to obtain a handle on the statistical significance of our results.

This paper is structured as follows. 
In Section~\ref{sec:numerical} we describe the most relevant details of
the simulations, and in Section~\ref{sec:method} we discuss the method
we use to measure the DM halo shape. 
In Section~\ref{sec:results} we present our results focusing on the
radial shape trends at $z=0$, and the alignments with the stellar disc.
We place our findings in Section~\ref{sec:discussion} 
into the context of other numerical work, explore correlations of the
shape with baryonic properties in the disc, and finally make a direct
comparison against observational constraints for the MW's dark matter halo shape.
We give our conclusions in Section~\ref{sec:conclusions}.

\section{Numerical Simulations}
\label{sec:numerical}

The Auriga project carries out cosmological zoom-in simulations of MW-sized 
halos in a $\Lambda$CDM cosmology. 
The simulations come in two versions: dark matter only and full 
baryonic physics included (this entails magetohydrodynamics, gas cooling, star formation, black hole growth, and associated feedback processes).
A detailed description of the simulations, the numerical treatment, and the resulting galaxy properties can be found in \citet{auriga}.
Here we summarize some of the main features.

The simulated objects were chosen as a set of 30
isolated halos in the dark matter parent simulation of the Evolution and Assembly of GaLaxies and their
Environments (EAGLE)  project \citep{Eagle}.   
These halos were randomly selected from a sample of the most isolated
halos at $z=0$ whose virial mass $M_{200}$ was between $10^{12}\, {\rm M}_\odot$ and
$2\times 10^{12}\, {\rm M}_\odot$. 
The cosmological parameters in the simulations correspond to
$\Omega_m=0.307$, $\Omega_b=0.048$, $\Omega_\Lambda=0.693$ and a
dimensionless Hubble parameter $h=0.6777$ \citep{2014A&A...571A..16P}

The selected halos were then re-simulated at much higher resolution by applying a
zoom-in technique with variable spatial resolution using the moving-mesh code AREPO  \citep{arepo, 2013MNRAS.432..176P},
and by using a comprehensive model for galaxy formation physics
that includes gravity, ideal magnetohydrodynamics,  a phenomenological
description of star formation, chemical enrichment from supernovae
and stellar feedback.   
The simulations also follow the formation and evolution of black holes
together with  Active Galactic Nuclei (AGN) feedback

The 30 zoom-in halos have a dark matter particle mass of $\sim 3\times
10^5$\Msun while the baryonic mass resolution is $\sim 5\times 10^4$\Msun.
The gravitational force softening length for stellar
particles and high-resolution dark matter particles 
is fixed to be 500 $h^{-1}\,{\rm  pc}$ in comoving coordinates up to $z=1$,
and 396 pc in physical coordinates afterwards.
The gravitational softening length for gas cells changes with the mean
cell radius but is limited to be at least the stellar softening
length of 500 $h^{-1}$ pc comoving. 
 We refer to this setup as the ``Level-4'' resolution, based on the nomenclature of the Auriga project.
Six of these haloes were also simulated at a higher ``Level-3'' resolution, 
representing an increase by a factor of 8 in mass and a factor of 2 in spatial
resolution.  
We use the higher resolution haloes to define a minimal radius above
which our results can be considered converged.
From now on we refer to the haloes simulated with baryonic physics as the
MHD sample, and to the haloes simulated with dark matter only as the
DMO sample.

\begin{figure*}
\includegraphics[width=0.9\textwidth]{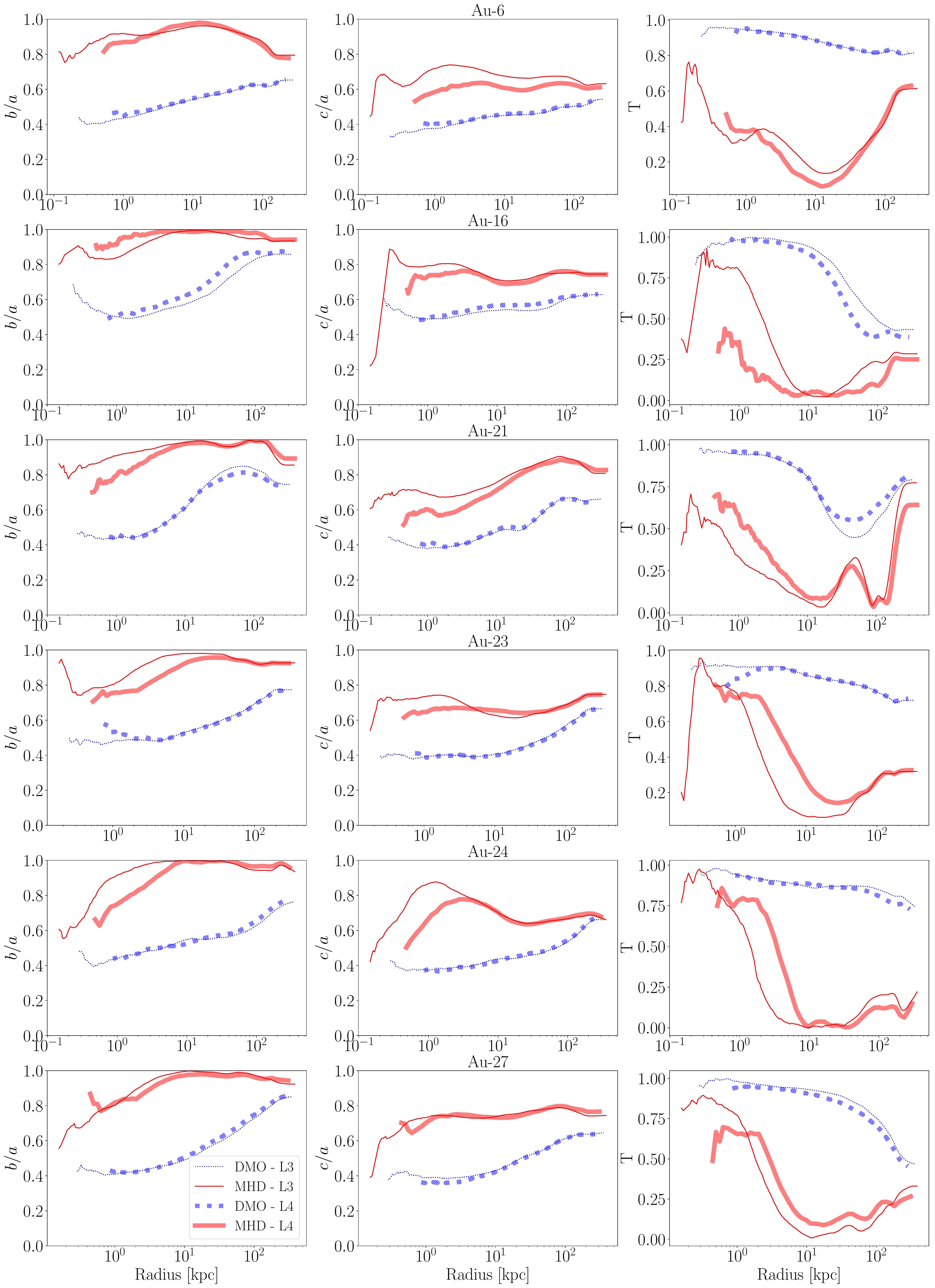}
\caption{Axes ratios for the six halos simulated at two different
  resolutions with two different physical models (DMO and MHD). 
  In DMO simulations there is good agreement between the two
  resolutions (L3 and L4) at all radii down to scales of $\approx 1$
  kpc from the center. 
  In MHD simulations, there are noticeable differences at scales below
  $\approx 10$ kpc. 
  In this paper, we only report results at scales larger than $\approx 16$ kpc
  ($R_{200}/16$).}
\label{fig:resolution}
\end{figure*}

\begin{table*}
  \centering 
  \begin{tabular}{c c c c c c c c c c c}
  \hline\hline
  Halo & $(b/a)_{16}$ & $(c/a)_{16}$& $(b/a)_{8}$ & $(c/a)_{8}$& $(b/a)_{4}$ & $(c/a)_{4}$& $(b/a)_{2}$ & $(c/a)_{2}$& $(b/a)_{1}$ & $(c/a)_{1}$ \\
  \hline
Au-1 & 0.48 & 0.41 & 0.52 & 0.44 & 0.59 & 0.48 & 0.66 & 0.53 & 0.72 & 0.55\\
Au-2 & 0.66 & 0.47 & 0.72 & 0.49 & 0.76 & 0.51 & 0.80 & 0.54 & 0.79 & 0.55\\
Au-3 & 0.56 & 0.42 & 0.61 & 0.47 & 0.70 & 0.55 & 0.79 & 0.65 & 0.85 & 0.72\\
Au-4 & 0.42 & 0.35 & 0.47 & 0.40 & 0.49 & 0.43 & 0.55 & 0.49 & 0.59 & 0.52\\
Au-5 & 0.49 & 0.41 & 0.53 & 0.47 & 0.56 & 0.51 & 0.59 & 0.56 & 0.66 & 0.59\\
Au-6 & 0.56 & 0.46 & 0.58 & 0.46 & 0.61 & 0.49 & 0.62 & 0.51 & 0.65 & 0.54\\
Au-7 & 0.46 & 0.37 & 0.45 & 0.37 & 0.51 & 0.40 & 0.55 & 0.41 & 0.62 & 0.46\\
Au-8 & 0.75 & 0.47 & 0.84 & 0.53 & 0.85 & 0.58 & 0.76 & 0.54 & 0.82 & 0.55\\
Au-9 & 0.49 & 0.40 & 0.52 & 0.43 & 0.54 & 0.45 & 0.60 & 0.52 & 0.65 & 0.60\\
Au-10 & 0.62 & 0.51 & 0.61 & 0.50 & 0.58 & 0.49 & 0.60 & 0.53 & 0.65 & 0.59\\
Au-11 & 0.47 & 0.37 & 0.36 & 0.29 & 0.32 & 0.28 & 0.41 & 0.37 & 0.47 & 0.45\\
Au-12 & 0.66 & 0.53 & 0.76 & 0.59 & 0.83 & 0.66 & 0.88 & 0.68 & 0.94 & 0.71\\
Au-13 & 0.60 & 0.52 & 0.61 & 0.53 & 0.64 & 0.54 & 0.70 & 0.56 & 0.73 & 0.54\\
Au-14 & 0.97 & 0.67 & 0.95 & 0.66 & 0.90 & 0.64 & 0.89 & 0.69 & 0.87 & 0.75\\
Au-15 & 0.61 & 0.42 & 0.65 & 0.48 & 0.68 & 0.52 & 0.66 & 0.50 & 0.70 & 0.54\\
Au-16 & 0.64 & 0.57 & 0.73 & 0.57 & 0.84 & 0.58 & 0.87 & 0.62 & 0.88 & 0.63\\
Au-17 & 0.98 & 0.79 & 0.98 & 0.83 & 0.95 & 0.84 & 0.91 & 0.81 & 0.93 & 0.87\\
Au-18 & 0.56 & 0.55 & 0.57 & 0.55 & 0.60 & 0.57 & 0.64 & 0.59 & 0.62 & 0.56\\
Au-19 & 0.58 & 0.51 & 0.59 & 0.50 & 0.65 & 0.54 & 0.73 & 0.61 & 0.80 & 0.70\\
Au-20 & 0.61 & 0.44 & 0.68 & 0.48 & 0.72 & 0.51 & 0.76 & 0.59 & 0.81 & 0.70\\
Au-21 & 0.67 & 0.50 & 0.75 & 0.50 & 0.81 & 0.61 & 0.80 & 0.66 & 0.74 & 0.64\\
Au-22 & 0.68 & 0.56 & 0.71 & 0.59 & 0.77 & 0.63 & 0.82 & 0.71 & 0.83 & 0.78\\
Au-23 & 0.54 & 0.41 & 0.58 & 0.44 & 0.62 & 0.49 & 0.70 & 0.57 & 0.77 & 0.66\\
Au-24 & 0.54 & 0.44 & 0.57 & 0.47 & 0.59 & 0.49 & 0.66 & 0.55 & 0.75 & 0.66\\
Au-25 & 0.59 & 0.56 & 0.66 & 0.65 & 0.71 & 0.66 & 0.62 & 0.55 & 0.69 & 0.61\\
Au-26 & 0.52 & 0.41 & 0.58 & 0.43 & 0.65 & 0.43 & 0.73 & 0.47 & 0.75 & 0.51\\
Au-27 & 0.53 & 0.45 & 0.61 & 0.51 & 0.68 & 0.56 & 0.76 & 0.62 & 0.84 & 0.64\\
Au-28 & 0.49 & 0.42 & 0.49 & 0.44 & 0.56 & 0.48 & 0.64 & 0.52 & 0.74 & 0.58\\
Au-29 & 0.50 & 0.40 & 0.56 & 0.45 & 0.61 & 0.48 & 0.69 & 0.54 & 0.72 & 0.60\\
Au-30 & 0.62 & 0.44 & 0.77 & 0.54 & 0.87 & 0.58 & 0.85 & 0.58 & 0.75 & 0.53\\
  \hline\hline
  \end{tabular}
  \caption{Axes ratios of Auriga dark matter halos at $z=0$ at
    different radii in the DMO simulations. The subscripts $16$, $8$, $4$, $2$, and $1$,
    indicate that the shape was measured at radii of $R_{200}/16$,
    $R_{200}/8$, $R_{200}/4$, $R_{200}/2$, and $R_{200}$, respectively.   \label{table:DMO}}
\end{table*}

\begin{table*}
  \centering 
  \begin{tabular}{c c c c c c c c c c c}
  \hline\hline
  Halo & $(b/a)_{16}$ & $(c/a)_{16}$& $(b/a)_{8}$ & $(c/a)_{8}$& $(b/a)_{4}$ & $(c/a)_{4}$& $(b/a)_{2}$ & $(c/a)_{2}$& $(b/a)_{1}$ & $(c/a)_{1}$ \\
  \hline
Au-1 & 0.94 & 0.65 & 0.96 & 0.66 & 0.96 & 0.70 & 0.92 & 0.68 & 0.91 & 0.67\\
Au-2 & 0.93 & 0.61 & 0.95 & 0.58 & 0.96 & 0.61 & 0.96 & 0.65 & 0.88 & 0.65\\
Au-3 & 0.97 & 0.66 & 0.96 & 0.66 & 0.94 & 0.70 & 0.93 & 0.76 & 0.94 & 0.79\\
Au-4 & 0.88 & 0.77 & 0.82 & 0.79 & 0.80 & 0.76 & 0.85 & 0.79 & 0.83 & 0.76\\
Au-5 & 0.97 & 0.79 & 0.94 & 0.80 & 0.92 & 0.81 & 0.94 & 0.82 & 0.94 & 0.81\\
Au-6 & 0.98 & 0.60 & 0.96 & 0.60 & 0.92 & 0.63 & 0.86 & 0.63 & 0.78 & 0.61\\
Au-7 & 0.95 & 0.66 & 0.95 & 0.66 & 0.97 & 0.68 & 0.97 & 0.69 & 0.96 & 0.70\\
Au-8 & 0.99 & 0.59 & 0.97 & 0.62 & 0.92 & 0.66 & 0.76 & 0.57 & 0.83 & 0.58\\
Au-9 & 0.94 & 0.75 & 0.93 & 0.73 & 0.92 & 0.73 & 0.93 & 0.79 & 0.92 & 0.82\\
Au-10 & 0.95 & 0.85 & 0.94 & 0.85 & 0.93 & 0.85 & 0.92 & 0.82 & 0.91 & 0.80\\
Au-11 & 0.54 & 0.49 & 0.55 & 0.48 & 0.61 & 0.53 & 0.60 & 0.53 & 0.62 & 0.57\\
Au-12 & 0.95 & 0.77 & 0.88 & 0.81 & 0.90 & 0.84 & 0.92 & 0.81 & 0.95 & 0.82\\
Au-13 & 0.95 & 0.88 & 0.96 & 0.88 & 0.98 & 0.82 & 0.94 & 0.76 & 0.82 & 0.62\\
Au-14 & 0.95 & 0.75 & 0.93 & 0.79 & 0.91 & 0.80 & 0.91 & 0.79 & 0.93 & 0.81\\
Au-15 & 0.99 & 0.67 & 0.98 & 0.75 & 0.94 & 0.79 & 0.93 & 0.81 & 0.88 & 0.76\\
Au-16 & 0.99 & 0.69 & 0.99 & 0.70 & 0.99 & 0.74 & 0.98 & 0.76 & 0.99 & 0.99\\
Au-17 & 0.96 & 0.79 & 0.96 & 0.80 & 0.97 & 0.82 & 0.96 & 0.83 & 0.96 & 0.87\\
Au-18 & 0.93 & 0.76 & 0.92 & 0.74 & 0.86 & 0.72 & 0.81 & 0.71 & 0.74 & 0.66\\
Au-19 & 0.95 & 0.67 & 0.91 & 0.63 & 0.89 & 0.63 & 0.83 & 0.64 & 0.85 & 0.69\\
Au-20 & 0.81 & 0.58 & 0.91 & 0.64 & 0.95 & 0.71 & 0.97 & 0.78 & 0.86 & 0.77\\
Au-21 & 0.98 & 0.74 & 0.97 & 0.82 & 0.97 & 0.87 & 0.99 & 0.88 & 0.99 & 0.99\\
Au-22 & 0.96 & 0.84 & 0.93 & 0.84 & 0.93 & 0.85 & 0.95 & 0.87 & 0.97 & 0.88\\
Au-23 & 0.94 & 0.65 & 0.96 & 0.64 & 0.94 & 0.65 & 0.92 & 0.69 & 0.93 & 0.74\\
Au-24 & 0.99 & 0.67 & 1.00 & 0.63 & 0.99 & 0.65 & 0.97 & 0.68 & 0.98 & 0.70\\
Au-25 & 0.97 & 0.71 & 0.96 & 0.73 & 0.86 & 0.74 & 0.67 & 0.64 & 0.71 & 0.67\\
Au-26 & 0.96 & 0.76 & 0.97 & 0.73 & 0.97 & 0.70 & 0.99 & 0.68 & 0.97 & 0.67\\
Au-27 & 0.98 & 0.73 & 0.97 & 0.75 & 0.97 & 0.79 & 0.96 & 0.79 & 0.95 & 0.77\\
Au-28 & 0.98 & 0.82 & 0.94 & 0.77 & 0.88 & 0.73 & 0.86 & 0.74 & 0.93 & 0.75\\
Au-29 & 0.91 & 0.81 & 0.89 & 0.72 & 0.87 & 0.69 & 0.84 & 0.68 & 0.85 & 0.69\\
Au-30 & 0.89 & 0.73 & 0.78 & 0.64 & 0.69 & 0.53 & 0.80 & 0.58 & 0.91 & 0.63\\
  \hline\hline
  \end{tabular}
  \caption{Same as Table~\ref{table:DMO}, but for the MHD simulations.   \label{table:MHD}}
\end{table*}

\begin{figure*}
\includegraphics[width=0.45\textwidth]{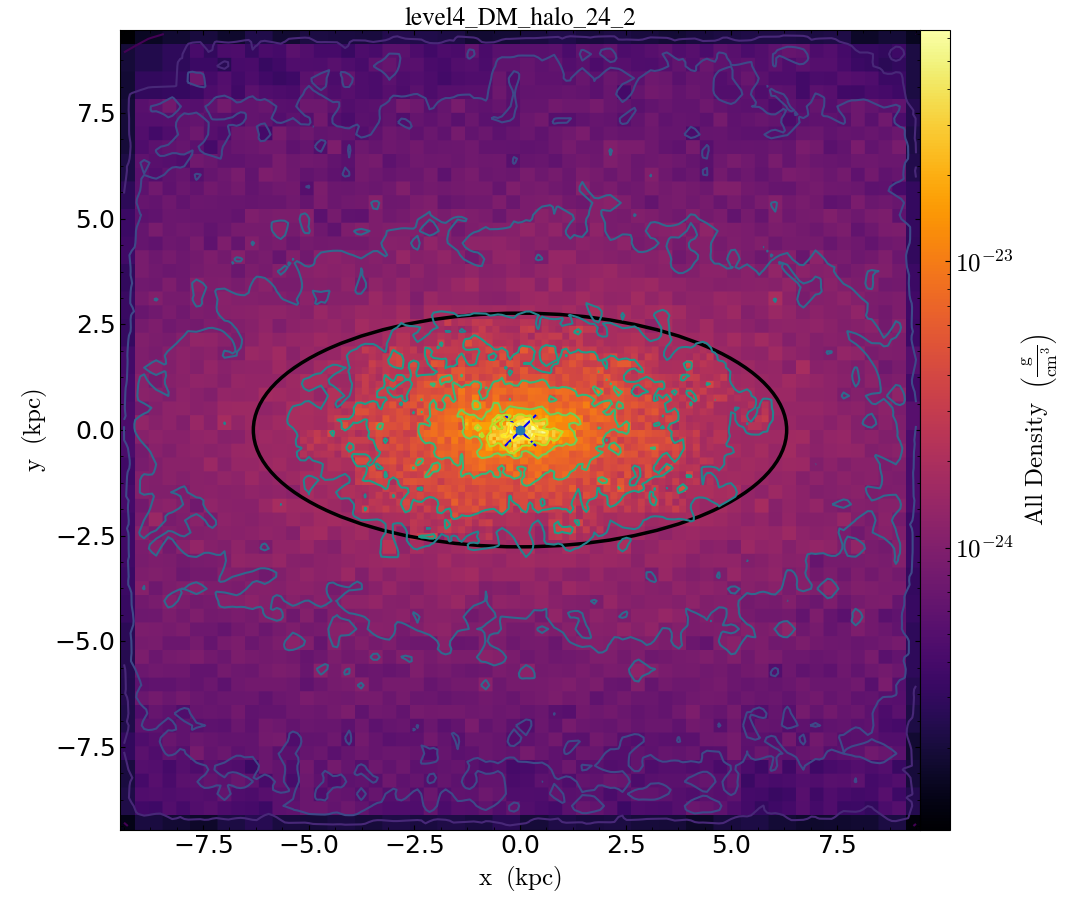}
\includegraphics[width=0.45\textwidth]{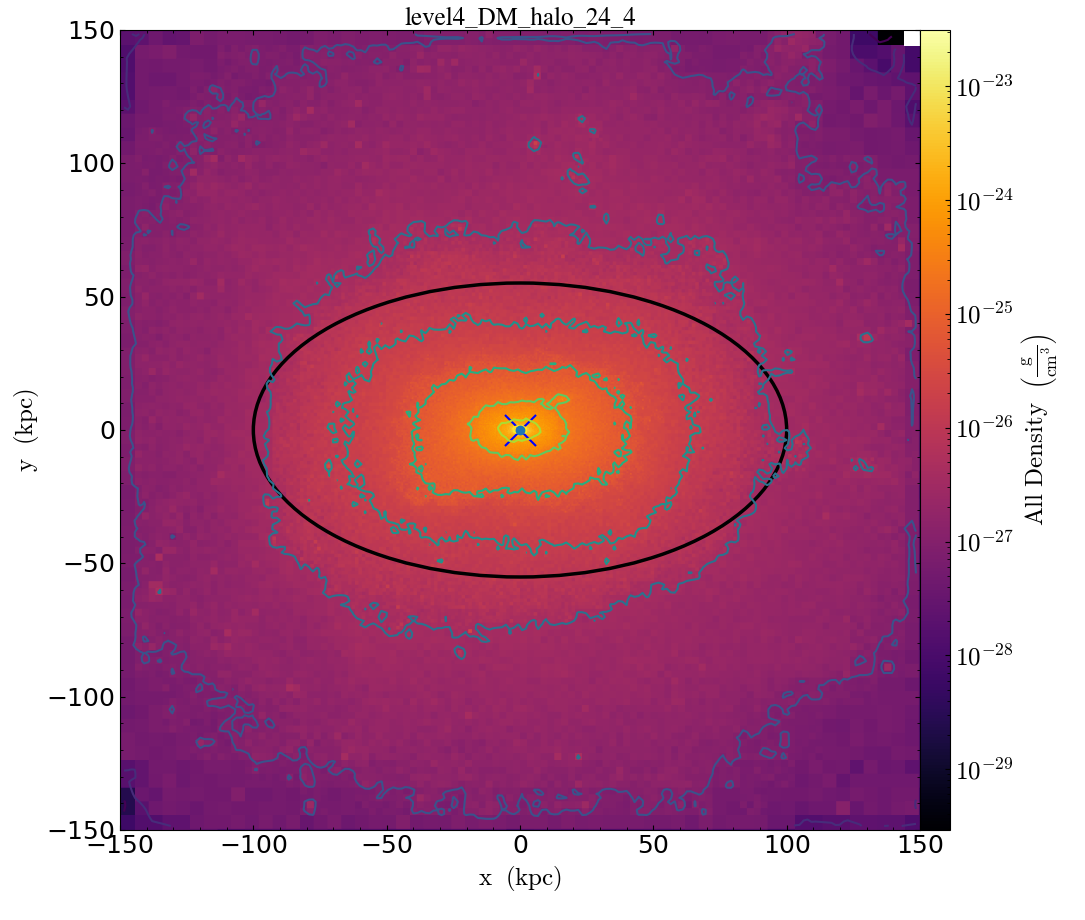}
\includegraphics[width=0.45\textwidth]{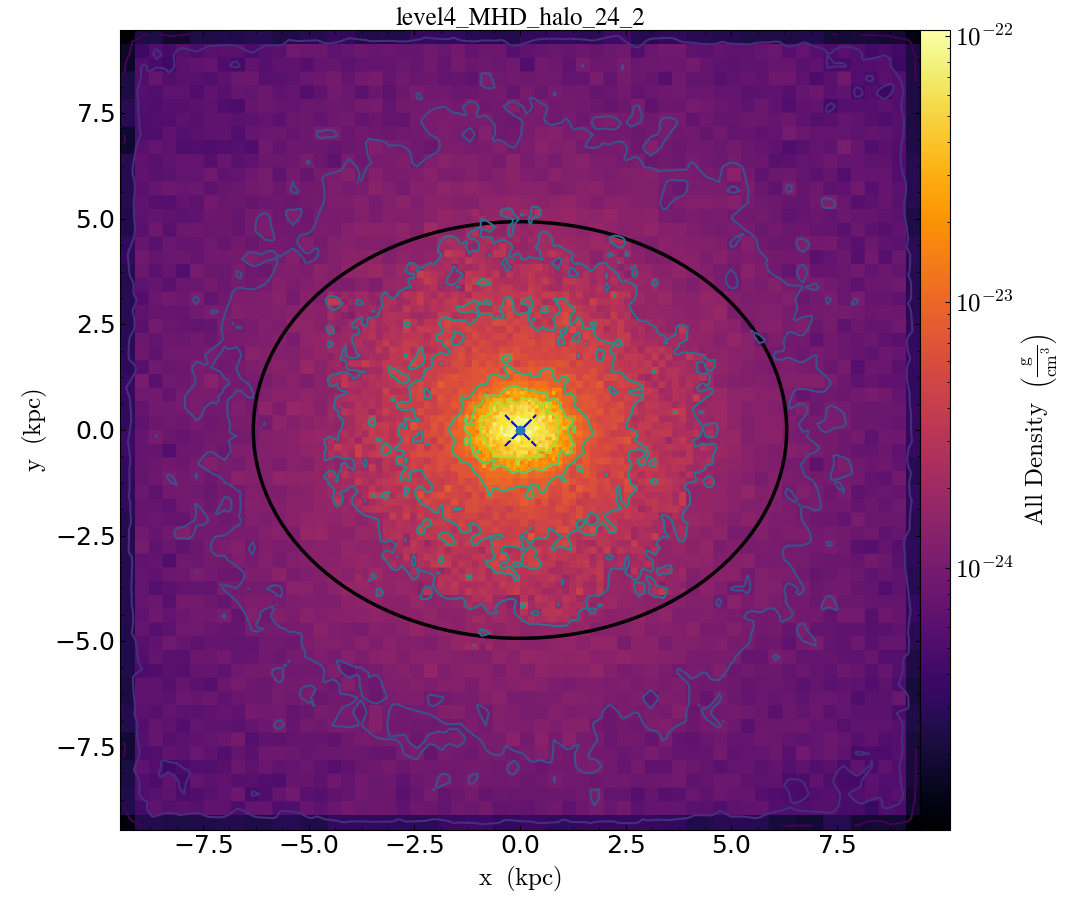}
\includegraphics[width=0.45\textwidth]{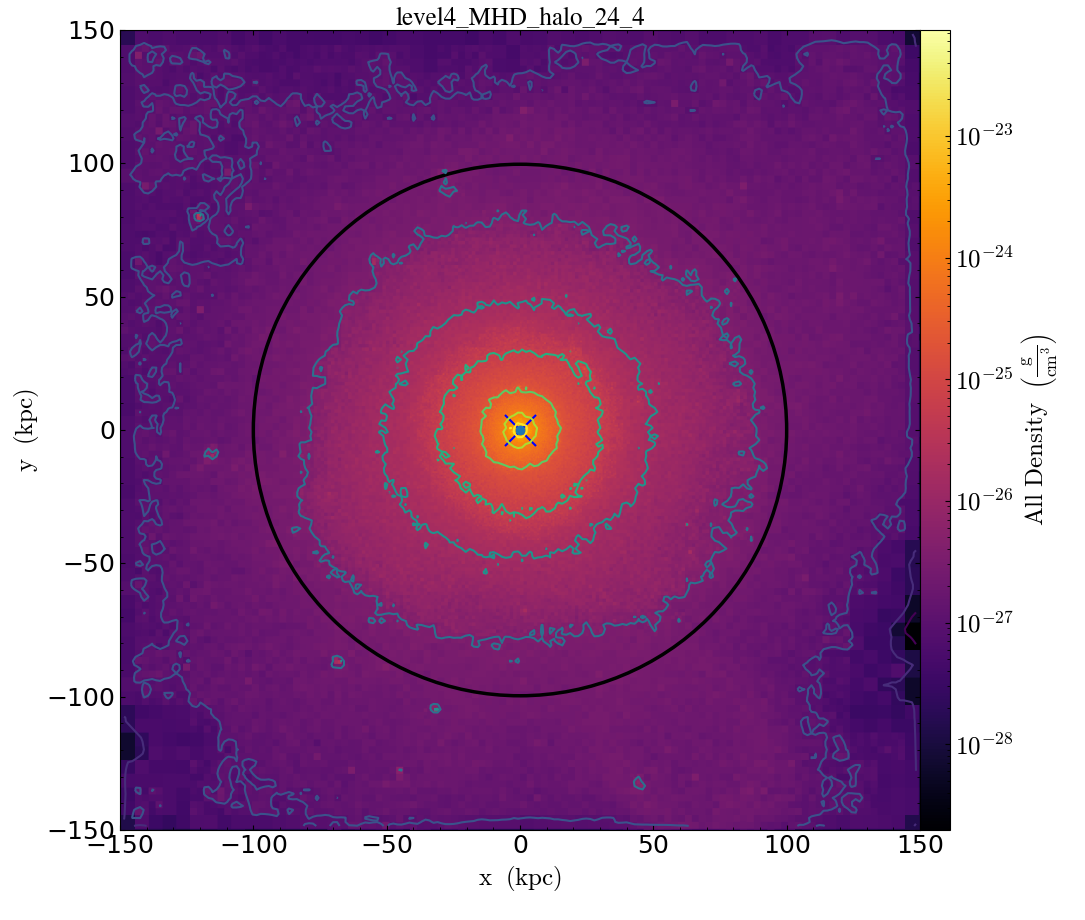}
  \caption{DM density on a logarithmic color scale within a slice of thickness one tenth
    of the virial radius.
    The cut is perpendicular to the short axis of the inertia tensor ellipsoid.
    The black ellipses show the results of the fitting procedure
    described in Section~\ref{sec:method}. 
    Upper/lower panels correspond to DMO/MHD simulations, respectively.
    Left/right panels show data at small/large radii, respectively.
    This plot showcases the most noticeable effect in all halos
    across the Auriga simulations: haloes become rounder at all radii
    once baryonic physics is included.}
\label{fig:slices}
\end{figure*}

\begin{figure*}
\begin{center}
\includegraphics[width=0.8\columnwidth]{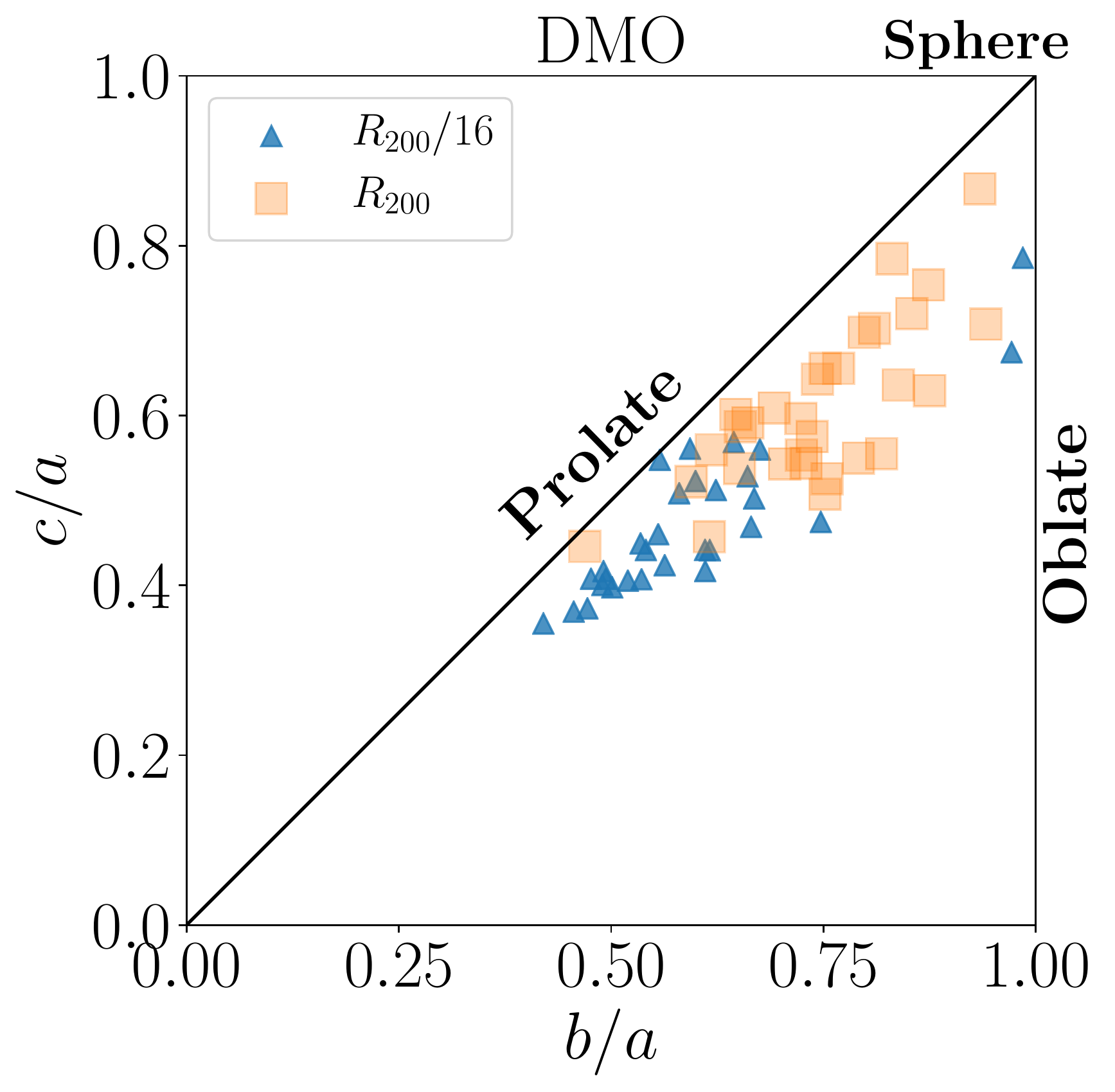}
\includegraphics[width=0.8\columnwidth]{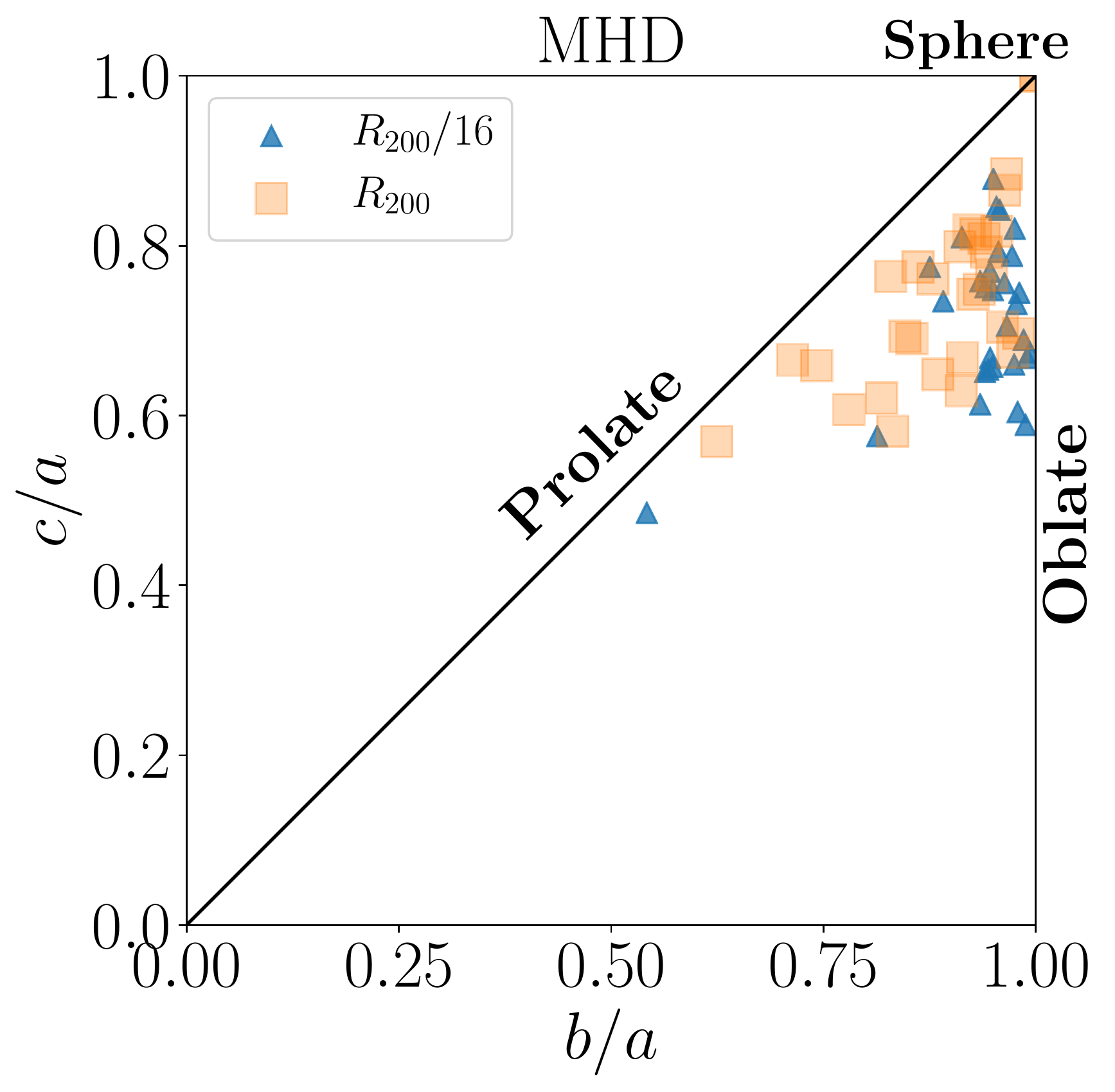}
\end{center}
\caption{Axes ratios for all simulated halos.
  Left/right panel corresponds to DMO/MHD simulations, respectively.
  Triangles (squares) represent the measurements at $R_{200}/16$
  ($R_{200}$) that correspond to physical distances of $\approx14$ kpc
  ($\approx 230$ kpc).
  Here we visualize three main population trends.
  First, in DMO simulations halos are rounder in the outskirts
  than in the inner part.
  Second, halos in MHD are rounder in the inner regions than in
  the outskirts (opposite to the DMO trend). 
  Third, halos in the MHD simulations are rounder than their DMO counterparts.
  These three points are further clarified in
  Figure~\ref{fig:triaxial_cumulative} through cumulative triaxiality distributions.}
  \label{fig:triaxiality_plane}
\end{figure*}

\begin{figure*}
\includegraphics[width=0.32\textwidth]{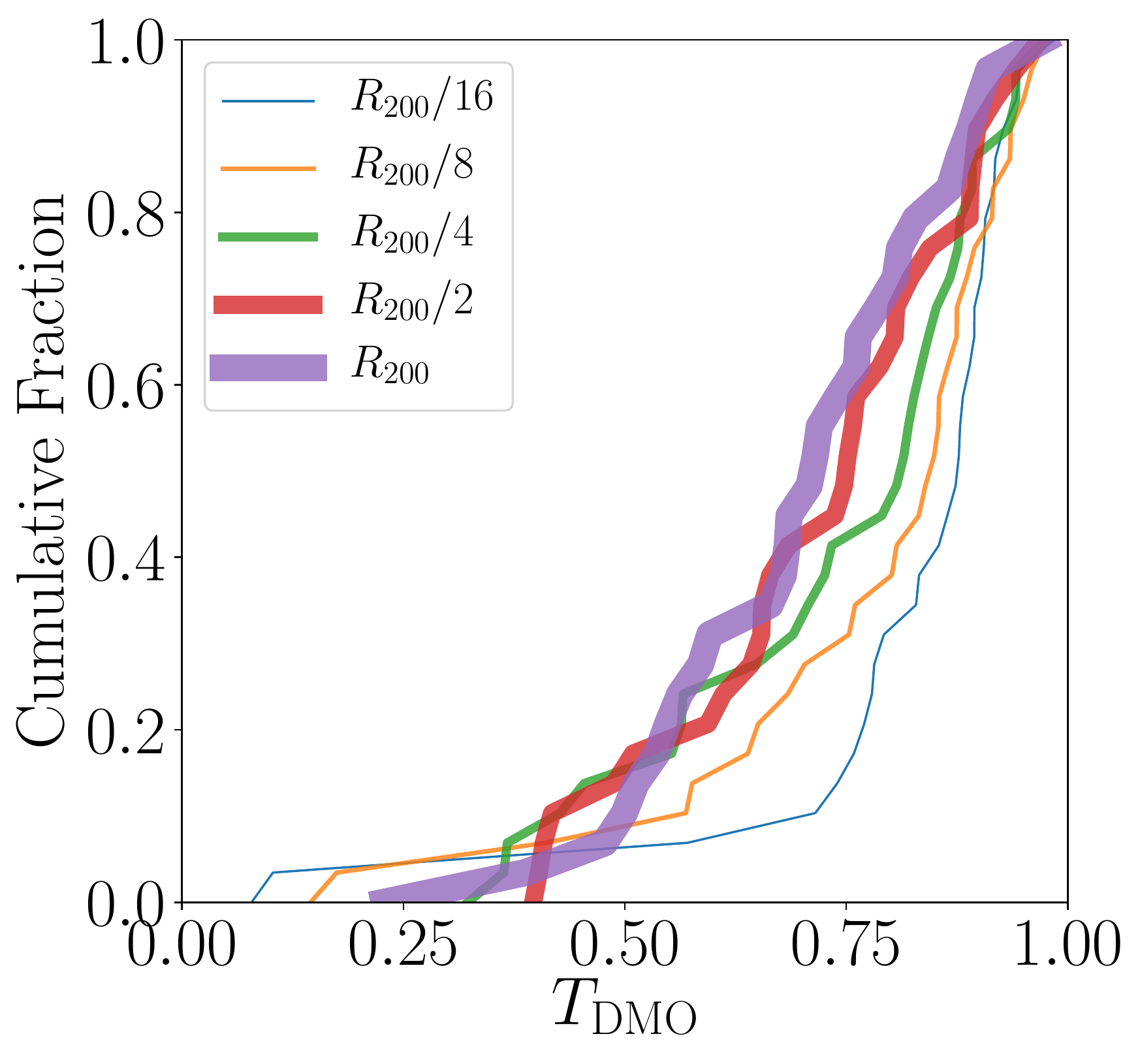}
\includegraphics[width=0.32\textwidth]{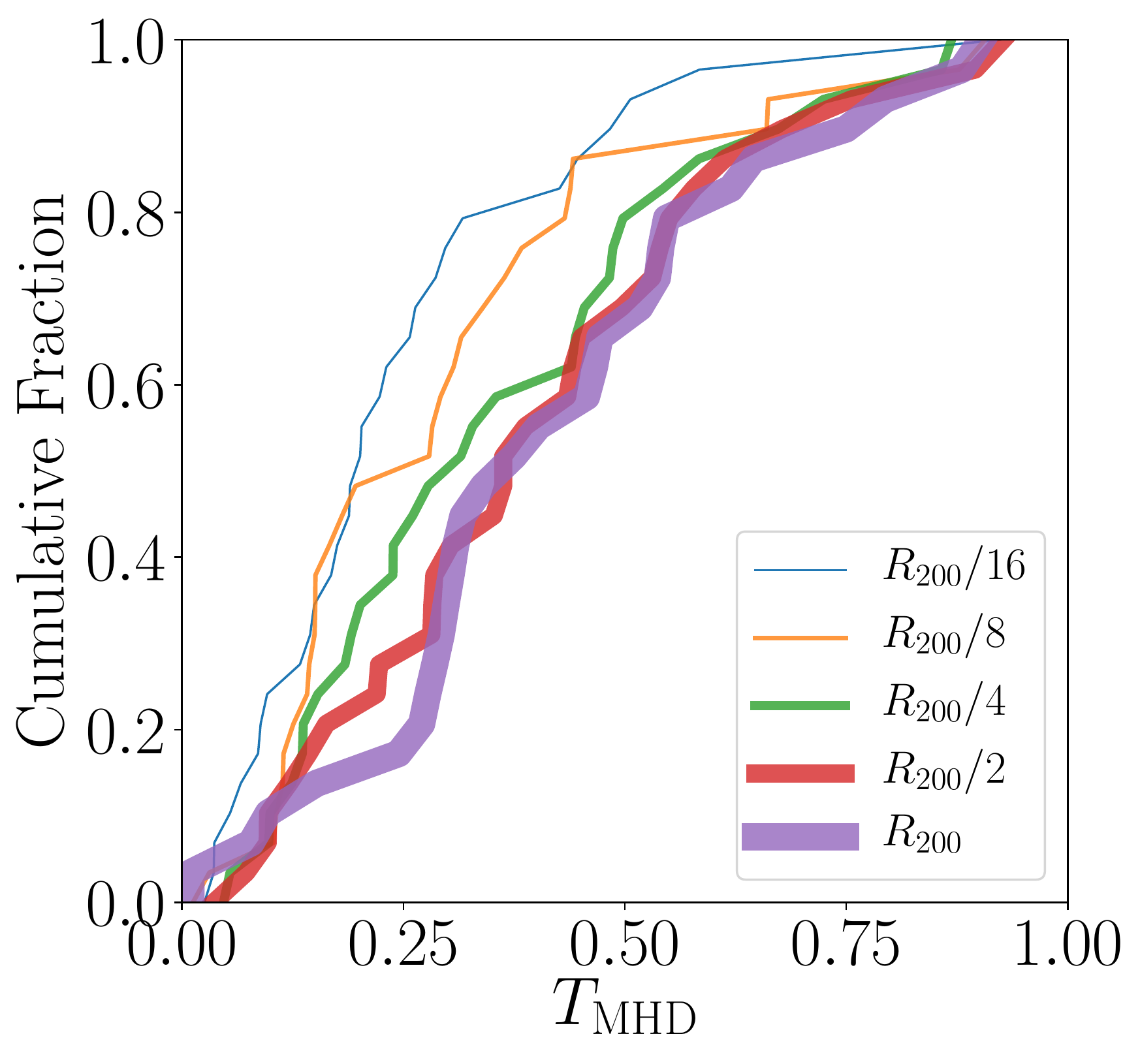}
\includegraphics[width=0.32\textwidth]{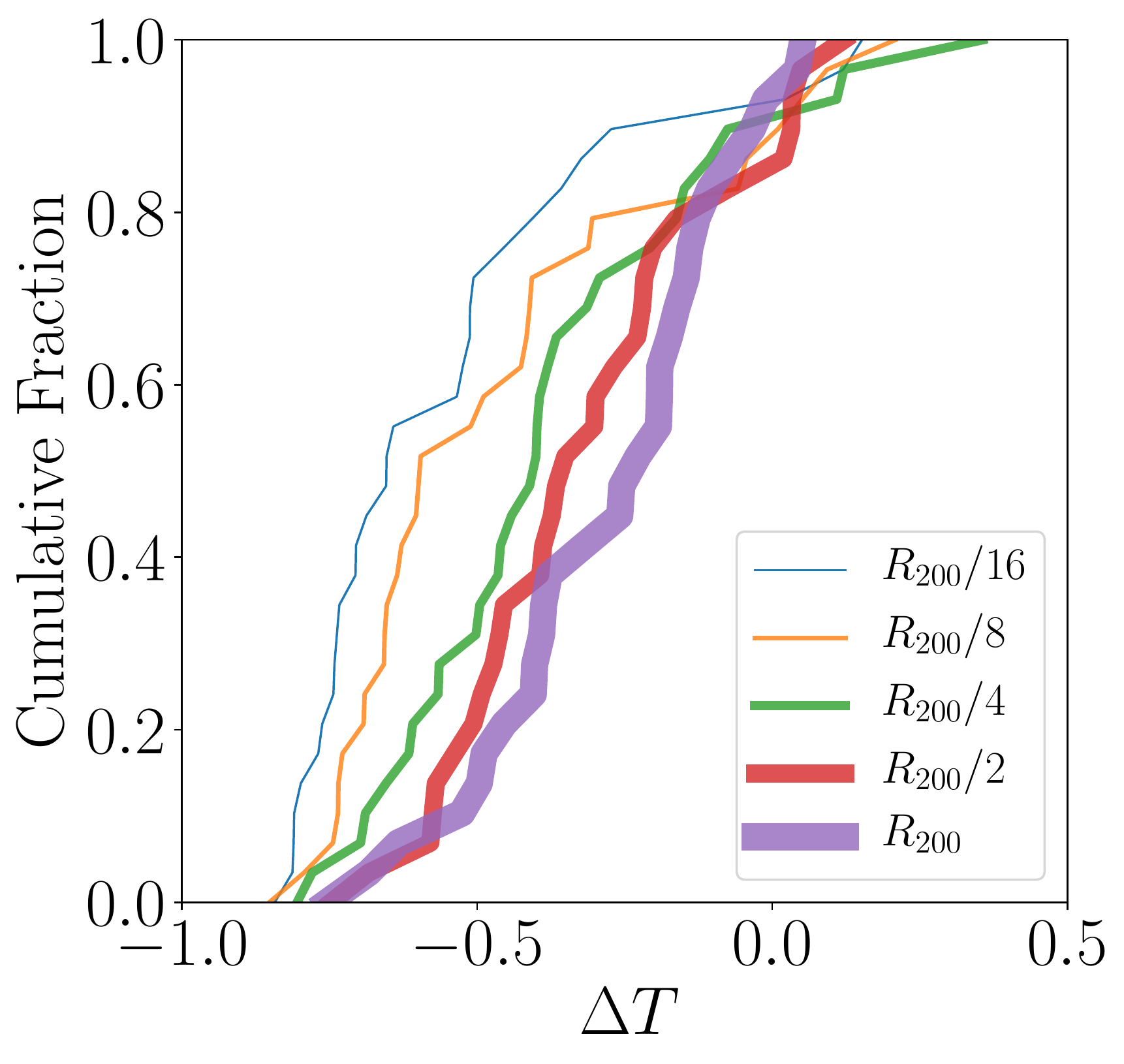}
\caption{Cumulative distributions of the triaxiality at five different radii.
  Right/middle panels correspond to the cumulative distribution of
  triaxiality in the DMO/MHD simulations, respectively. 
  In DMO simulations, the median triaxiality at all radii is larger
  than $2/3$. Furthermore, the triaxiality increases as one moves
  towards the inner part of the halo.
  In MHD simulations, this trend reverses.
  The median triaxiality at all radii is smaller than $1/3$, and the
  halo becomes less triaxial as on moves towards the stellar disc.
  The right panel shows the cumulative distribution of the changes
  in triaxiality in simulations with different physics, $\Delta
  T=T_{\rm MHD}-T_{\rm DMO}$. 
  At the virial radius, all halos become less triaxial in the MHD
  simulations, and the change is stronger towards the halo center.
  Only two haloes in the sample show the opposite trend of becoming
  more triaxial (but only at the smallest radii).} 
\label{fig:triaxial_cumulative}
\end{figure*}

\begin{figure*}
\begin{center}
\includegraphics[width=0.9\textwidth]{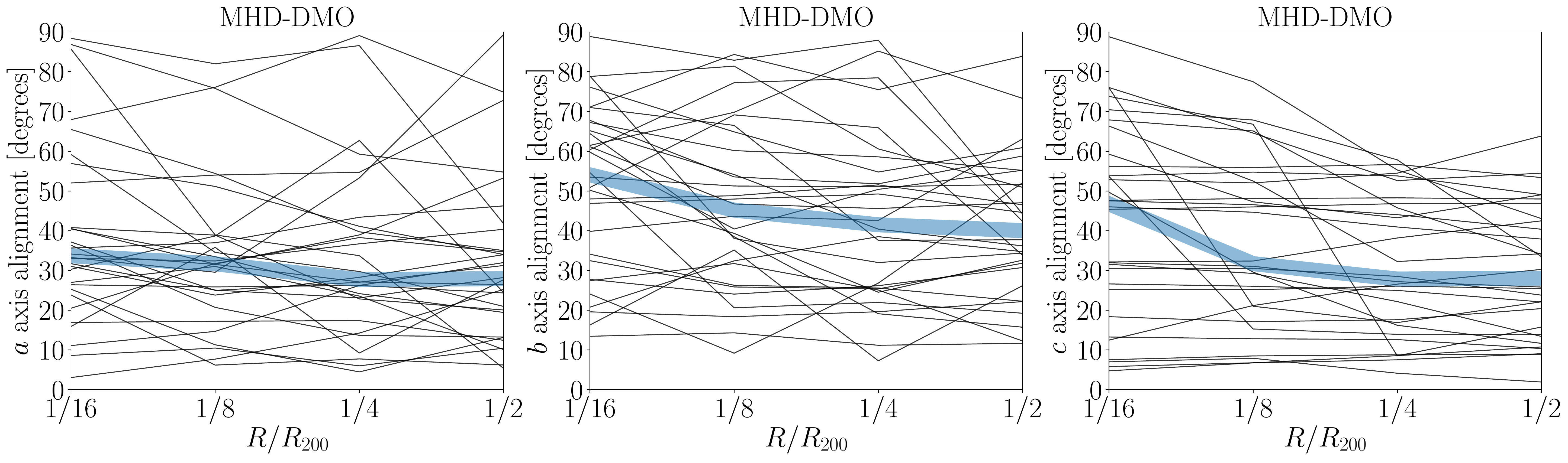}
\end{center}
\caption{Angles between the principal axis of the dark matter halo
  shape in the MHD and DMO simulations as a function of radius.
  Each panel compares the alignment of the corresponding
  major, middle, and minor axis in the halo.
  Thin lines correspond to each one of the thirty halos in the sample
  while the thick line traces the median value as a function of  radius.
 The halo shape in the MHD simulation is weakly correlated to the DMO case.}
\label{fig:angles_alignment}
\end{figure*}

\begin{figure*}
\begin{center}
\includegraphics[width=0.9\textwidth]{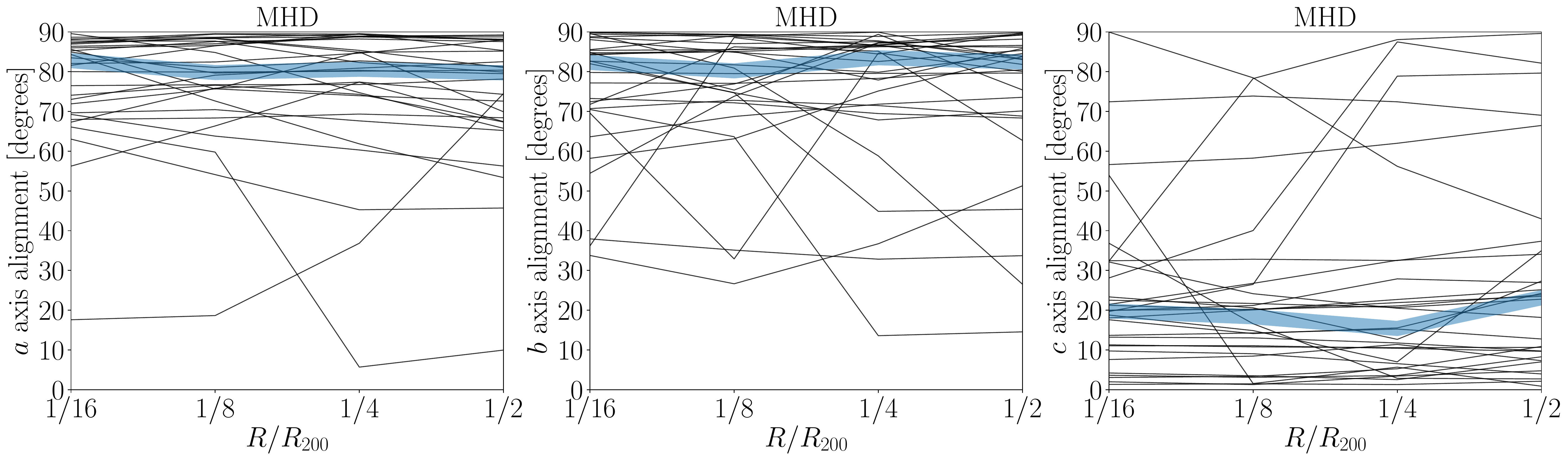}
\end{center}
\caption{Angles between the principal axis of the dark matter halo
  shape and the angular momentum of the stellar disc 
  as a function of the radius at
  which the halo shape directions are measured.
  Each panel compares the alignment of the corresponding
  major, middle, and minor axis in the halo.
  Thin lines correspond to each one of the thirty halos in the sample
  while the thick line traces the median value as a function of  radius.
  The sample presents a good alignment of the angular
  momentum with the minor halo axis. 
  This alignment is also constant in radius.
  However, the dark matter shells twist significantly in six halos 
  of our sample making the alignment change. 
}
\label{fig:cumulative_alignment}
\end{figure*}

\begin{figure*}
\begin{center}
\includegraphics[width=1.0\textwidth]{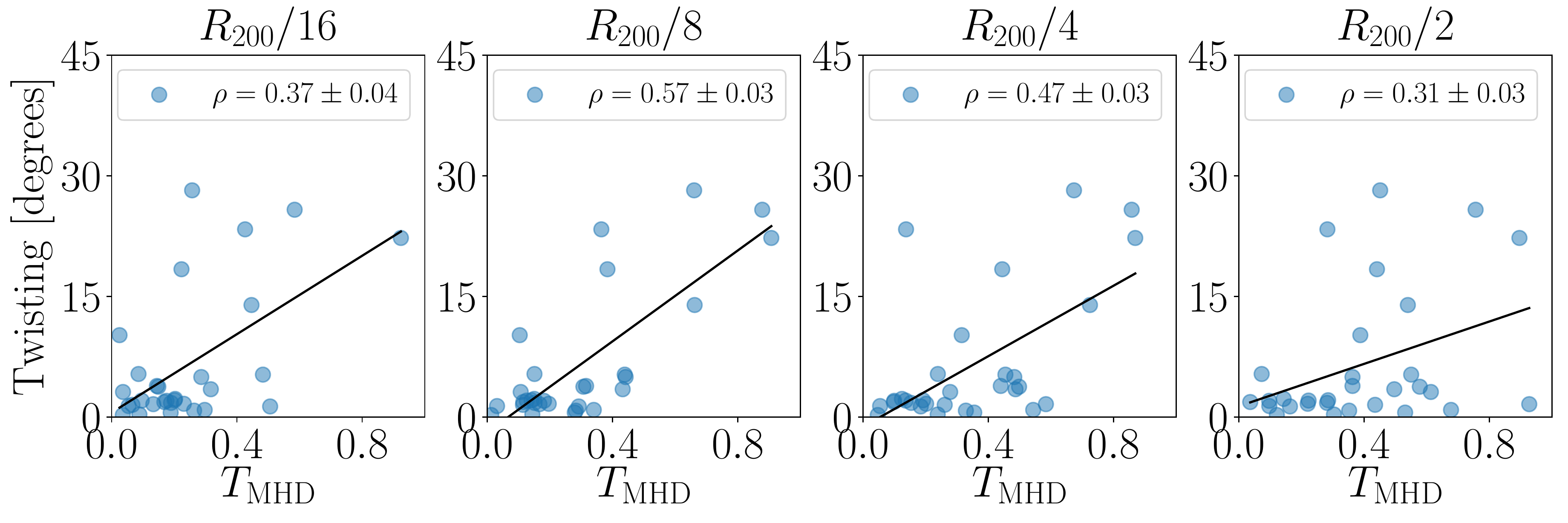}
\end{center}
\caption{Twisting in the halo-disc alignment (measured as the standard
  deviation of the alignment angles at four different radii)
  as a function of the halo triaxialiy at the radius indicated
  in each panel's title.
  The label with the $\rho$ value corresponds to the Spearman's rank
  correlation coefficient (mean value and uncertainty estimated via
  jackknife resampling) while the line shows the result of a
  minimum squares fit to the data.
  Twisting does not correlate with triaxiality in the same way at every
  radius. The strongest correlation appears around $0.12\, R_{200}$ ($28\pm2$ kpc).} 
\label{fig:alignment_correlations}
\end{figure*}

\begin{figure*}
\begin{center}
\includegraphics[width=0.8\textwidth]{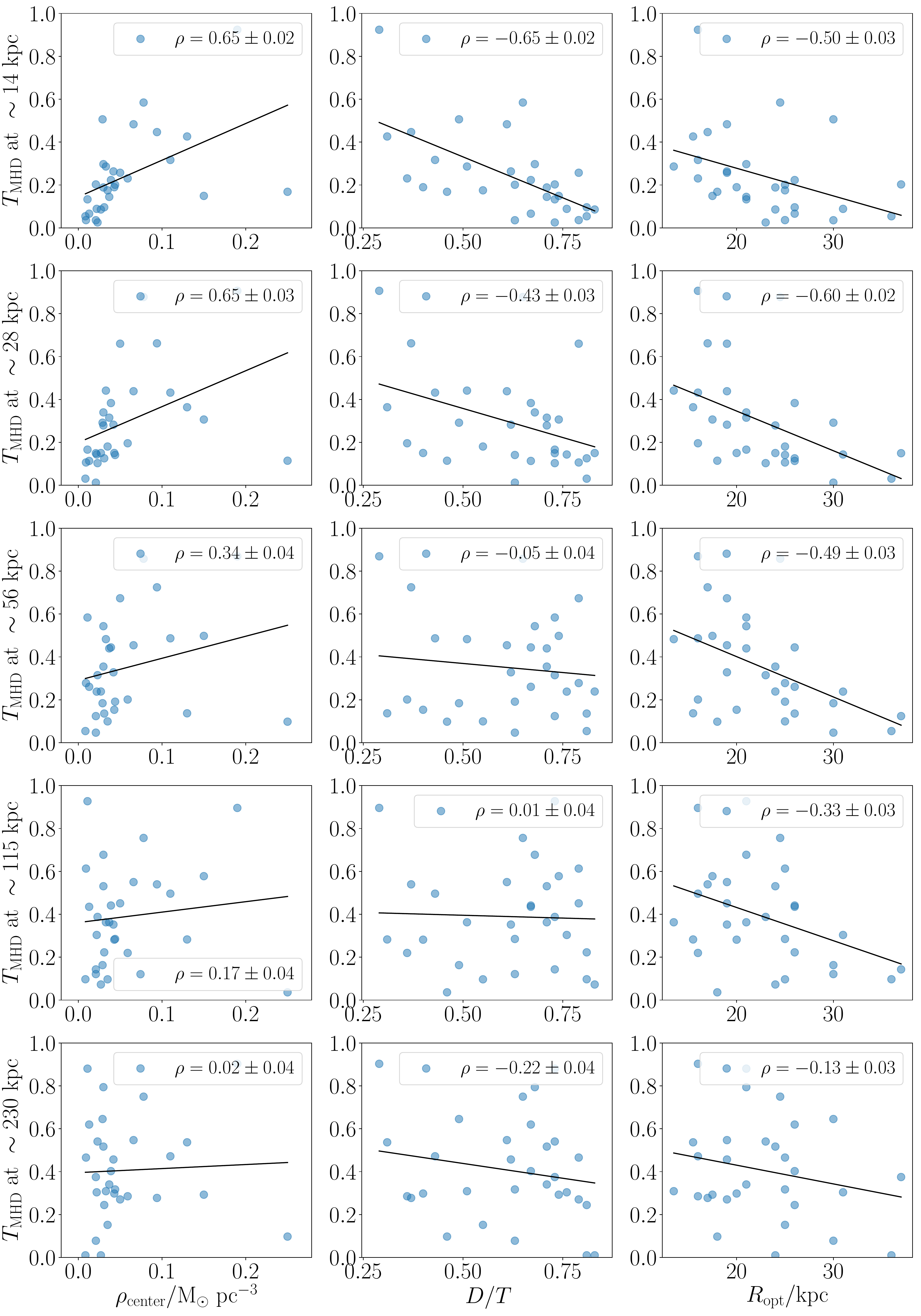}
\end{center}
\caption{Correlations between the halo triaxiality at different radii
  and baryonic disc properties. 
  The label with the $\rho$ value corresponds to the Spearman's rank
  correlation coefficient (mean value and uncertainty estimated via
  jackknife resampling).
  The line is the best minimum squares fit to a line.
  The $x$-axis in the first column is the gas density at the center of
  the galaxy within a sphere of radius  $1$ kpc \citep{Pakmor17};
  the second column shows the disc to total mass ratio, and the last
  column includes the disc optical radius defined to be the radius at which the
  $B$-band surface brightness drops below 25 mag arcsec$^{-2}$ \citep{auriga}.
  The largest correlations are found for the two smaller radii
  ($R_{200}/16$ and $R_{200}/8$).
  Extended and massive stellar discs with a low gas content at their
  cores are correlated with low dark matter triaxialities.
  The correlation decreases as one approaches larger radii.}
\label{fig:disc_correlations}
\end{figure*}

\begin{figure*}
\begin{center}
\includegraphics[width=0.9\textwidth]{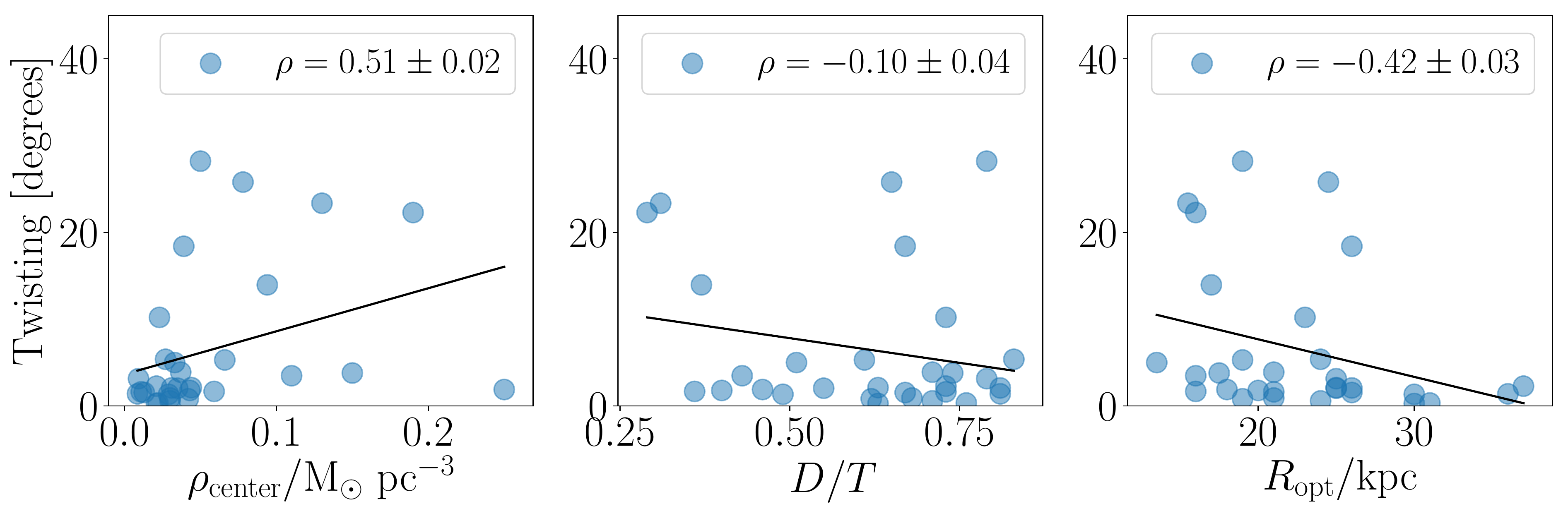}
\end{center}
\caption{Twisting in the halo-disc alingment as a function of the same
  baryonic disc properties as in Figure~\ref{fig:disc_correlations}.
  The label with the $\rho$ value corresponds to the Spearman's rank
  correlation coefficient (mean value and uncertainty estimated via
  jackknife resampling).
  The line is the best minimum squares fit to a line.
  In this case the disc to total mass ratio does not correlate with the alignment twisting.
  The gas density at the center and the disc optical radius still show
  a strong correlation, albeit weaker than the correlation of each with the triaxiality.}
\label{fig:twisting_correlations}
\end{figure*}

\begin{figure*}
\begin{center}
\includegraphics[width=0.9\textwidth]{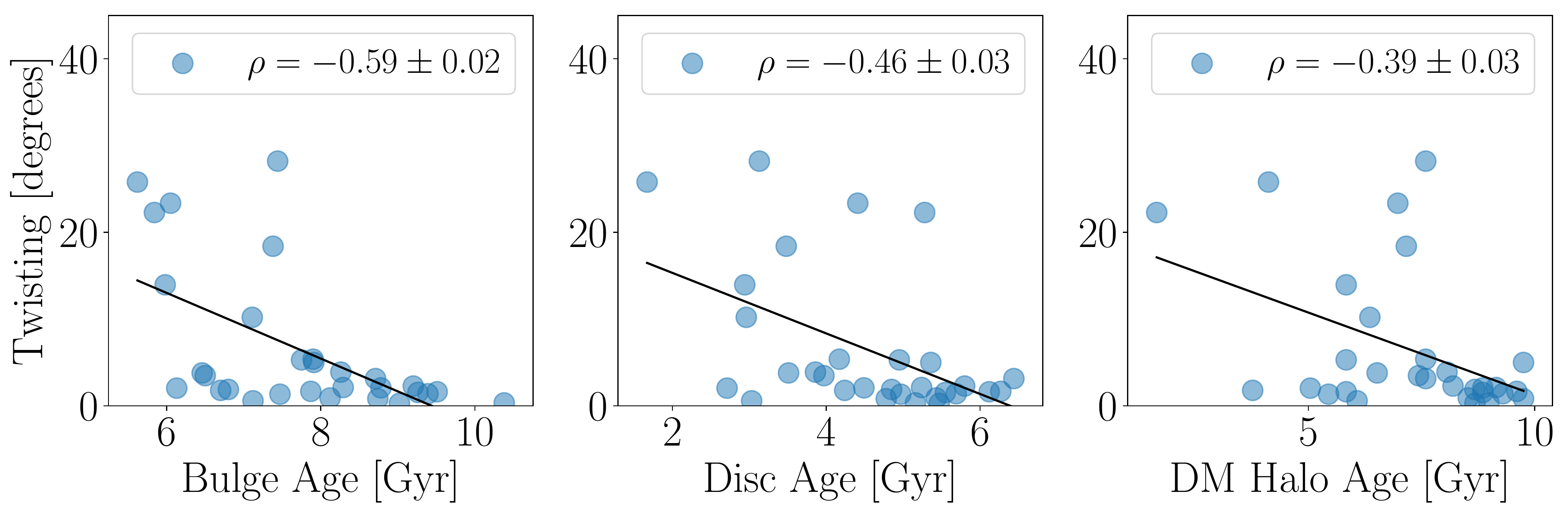}
\end{center}
\caption{Twisting in the halo-disc alingment as a function of the
  mean stellar age in the bulge/disc and the lookback formation time of the
  dark matter halo (time at which half of its present mass has assembled).
  The label with the $\rho$ value corresponds to the Spearman's rank
  correlation coefficient (mean value and uncertainty estimated via
  jackknife resampling).
  The line is the best minimum squares fit to a line.}
\label{fig:twisting_age}
\end{figure*}

\section{Halo Shape Measurement}
\label{sec:method}

The DM halo shape is usually estimated either based on the isopotential or
isodensity surfaces.   
Observational inferences typically target the 
isopotential contours which are probed by some dynamical tracers (gas, stars), while
simulations more often work with the isodensity contours which can be directly and easily
calculated from particle positions.  
Furthermore, it is well known that the density contours in thin shells are very sensitive to
the presence of small satellites \citep{Springel2004}.  
For this reason we measure the shape by taking
volume-enclosed particles, rather than shell-enclosed.  
This method yields results in good agreement to the isodensity
contours for radii $\leq 140$ kpc as explored by
\citet{VeraCiro11}.

In particular, we measure the shape using the reduced inertia tensor
\citep{Allgood06},  
\begin{equation}
I_{ij} = \sum_k \frac{x_k^{(i)}x_k^{(j)}}{d^2_k},
\label{eq:inertia}
\end{equation}
where the particle positions are measured with respect to the minimum of the
gravitational potential in each halo and are weighted by the inverse of 
the $k$-th particle distance squared, $d_k^2 = \sum_{i=1}^3 x_k^{(i)} x_k^{(i)}$, giving each particle the same
weight independent of distance.

Diagonalization of this tensor yields the eigenvectors and
eigenvalues that approximate the dark matter halo with an ellipsoidal shape.
The axis lenghts of this ellipsoid, $a\geq b \geq c$, are the square
roots of the  eigenvalues, and the directions of the principal
axes are the corresponding eigenvectors. 

We start the calculations taking into account all particles within a
sphere of radius $R_{\rm initial}$ and then iteratively recompute the
triaxial parameters by taking into account those particles within an ellipsoid of semi-axes
$r$, $r/q$ and $r/s$, and re-scaled distance $d^2=x^2+(y/q)^2+(z/s)^2$, where $q
= b/a$ and $s=c/a$ are the previously calculated axes ratios. 
We repeat this process until the average deviation of the semi-axes is
less than $10^{-6}$. For the converged shape, we define a single characteristic radius $R$ 
as the geometrical mean of the axes lengths $R=(abc)^{1/3}$.
We use this radial coordinate $R$ to parameterize the spatial changes
in halo shape we report in the following sections.
This is essentially the same method used to estimate the halo shape in the DM-only
Aquarius simulations \citep{VeraCiro11}. 

Following the convergence criterion by \cite{VeraCiro11} for DM-only simulations,
we restrict the measurement of the ellipsoidal parameters to radii
between $1$ kpc and $R_{200}$, where  $R_{200}$ corresponds to the viral radius
defined as the sphere  enclosing 200 times the critical density of the
Universe. However, the inner scale is similar to the gravitational smoothing length
for stellar particles in our default simulations.
For this reason, we use the level-3 simulations of Auriga performed at higher
resolution to gauge the convergence of the shape measurements in the
MHD simulations.

Figure \ref{fig:resolution} shows the results for the six halos
available both at level-3 and level-4 resolutions.
The dotted lines compare the DMO simulations showing that down to
scales of $1$ kpc the two resolutions yield very comparable results
\citep{VeraCiro11}. 

In MHD simulations the situation is less clean, as expected. 
The shape ratios $b/a$ and $c/a$ typically start to differ
from each other for different resolutions at radii smaller than $15$ kpc, 
with the exception of Au-27, which  shows good agreement of the shape 
measurements at all radii. 
For radii larger than $15$~kpc the shape ratios in the two resolutions have a difference of $0.01$ on average ($0.07$ in the worst case), while
for the triaxility the differences are $0.05$ on average 
($0.21$ in the worst case). 
For radii between $1$~kpc and $15$~kpc the shape ratios differ by $0.03$ on average
($0.21$ in the worst case) and the triaxility differences are $0.12$ on average ($0.54$ in the worst case). 
These average differences could be then considered as an uncertainty associated to numerical resolution on the results for each individual halo.
In order to have the smallest uncertainty associated to resolution, we concentrate in the following on measurements for radii larger  than $15$~kpc.

In our results we express the radius in units  of $R_{200}$ in
order to have a self-consistent dimensionless radial scale across all
halos. Given the similarity of the virial radii across halos, each dimensionless radius is
closely associated with a physical scale with a certain mean value and small standard deviation.
For instance, over the 30 halos in the sample, a dimensionless
radius of unity corresponds to a physical distance of $230\pm 16$
kpc.   Finally, we also measure the alignment of the halo shape against the
stellar disc angular momentum measured using the $10\%$ oldest stellar
particles belonging to the main Subfind structure within the virial radius.

\section{Results}
\label{sec:results}

Our main measurements for the halo shapes are summarized in Table \ref{table:DMO} for the
DMO simulations, and in Table \ref{table:MHD} for the MHD simulations.
These results are computed at five different radii, $R_{200}/16$,
$R_{200}/8$, $R_{200}/4$, $R_{200}/2$, and $R_{200}$.
In the following subsections, we discuss 
the results for the triaxiality, the halo-disc alignment and the
correlations with baryonic disc properties in greater detail.

\subsection{Triaxiality}

In the DMO sample, we find that halos become rounder with increasing radius.
The upper panels in Figure \ref{fig:slices} illustrate this effect.
The contours show a projected DM slice while the ellipsoid corresponds
to the full 3D shape determination. There we see a highly ellipsoidal
halo shape at radii $\approx 3$ kpc that becomes less elongated at
$\approx 50$ kpc. 

We summarize this trend in Figure \ref{fig:triaxiality_plane} by
plotting the results of all 30 halos of the DMO sample.
The left panel shows every halo in the $c/a$-$b/a$ plane at
two different radii, $R_{200}/16$ $(\approx 14$kpc$)$ and $R_{200}$. 
The outer part of the halo is systematically rounder than its inner
region. Nevertheless, the halo shape can still be considered to be prolate at
all radii. 

A different picture emerges for the MHD sample.
There all halos are rounder than their corresponding DMO
version at all radii. The lower panel of Figure~\ref{fig:slices} can be directly compared to
its MHD counterpart; there we observe how at large radii the hydrodynamic halo
becomes oblate and almost spherical. The right panel in Figure~\ref{fig:triaxiality_plane} shows the
results for the 30 halos in the MHD sample.

In Figure~\ref{fig:triaxial_cumulative}, we summarize the results at
different radii using the cumulative distributions for the 
triaxiality parameter $T$, defined as 
\begin{equation}
T=\frac{a^2-b^2}{a^2-c^2}.
\label{eq:triaxiality}
\end{equation}

The left panel of Figure \ref{fig:triaxial_cumulative} shows that in
the DMO sample the median triaxiality is larger than $2/3$ (a
typical value that marks the transition from triaxiality to
prolateness) at all radii. Furthermore this median value increases 
as we move towards the inner part of the halo.
The middle panel in the same Figure~\ref{fig:triaxial_cumulative}
shows how the trends in the MHD simulations go 
in the opposite direction with respect to the DMO results.
There the median triaxiality is always smaller than $1/3$ (a typical
value marking the transition from oblateness to triaxiality) and this
median value decreases as we move towards the inner part of the halo.

We now quantify this change in triaxiality at the level of individual halos.
We compute $\Delta T\equiv T_{\rm MHD}-T_{\rm DMO}$ to quantify the
change between the triaxiality in the MHD and the DMO simulations.
The right panel in Figure~\ref{fig:triaxial_cumulative} shows
the cumulative distribution at the same radii as in the other panels.
It is evident that at $R_{200}$ all the halos have
reduced their triaxiality after including baryonic physics.
At smaller radii this general trend continues, although 2 haloes
(Au-11 and Au-17)  in our sample increase their triaxiality with baryonic physics. 
However, this small fraction corresponds to halos that already were
outliers in the DMO sample and had triaxilities around $1/3$,
considerably lower than the mean of the parent sample.

Figure \ref{fig:angles_alignment} shows that the principal axis directions also
change in the MHD simulations with respect to the DMO counterpart. 
The major axis are aligned within $30\pm20$ degrees on average, while the median and minor
axes are aligned within $45\pm 20$ and $30\pm20$ degrees, respectively.
At $z=0$ the DM halo in the MHD simulations has a weak memory of the directions in the DMO setup.

\subsection{Halo-disc alignments}

A common assumption in dynamical models of the MW dark matter halo is that
its minor axis is perfectly aligned with the angular momentum of the
stellar disc. 
Although this is a reasonable assumption due to the stability of
the galactic disc in this configuration in simplified models of isolated galaxies, this
might not perfectly hold in a full cosmological context. 
To examine the degree of validity of this assumption we study in this
subsection the alignment between the stellar disc's angular momentum
eigenvectors and the dark matter halo shape.

Figure \ref{fig:cumulative_alignment} shows the
results as a function of radius.
The most striking feature of this plot is that most of the discs have
their angular momentum aligned with the minor axis of the halo,
perpendicular to the intermediate and major axis. 
The median and the standard deviation of the alignment angles with
the major/median/minor axis are $82\pm14$, $82\pm16$ and $19\pm20$
degrees, respectively.
This confirms statistically the expectation of a galactic disc aligned
with the minor axis of its dark matter halo, although some outliers
are present in the sample.

In six halos the alignment changes significantly as a
function of radius produced by the twisting in the shape ellipsoids. 
In contrast, the DMO simulations show coherent halo shapes as a
function of radius without twisting with the exception of a single halo. 
Because we are dealing with oblate halos $b/a\approx 1$, note that the
directions corresponding to the medium and major axis could be subject
to noise. 
For this reason, and in order to better quantify the twisting for each
halo, we take the standard deviation of the angle with the minor
axis, which should be less susceptible to noise.
With this measurement, twisting becomes a global property of the halo,
as the standard deviation is measured over all the different radii at
which we have quantified the alignment.

Figure~\ref{fig:alignment_correlations} shows the correlation between
the twisting and the corresponding triaxiality at the four different
radii at which the shape directions are measured.
The median, mean and standard deviation of the twisting 
are $2$, $6$ and $8$ degrees, respectively.
The correlation between twisting and triaxiality also changes with
radius as measured by the Spearman's rank correlation coefficient, a non-parametric way to measure the correlation between two variables; a perfect correlation would be given by a result of $+1$ or $-1$.
The largest correlation is obtained for the halo triaxiality measured at
$0.12R_{200}$. 
The most important consequence of this changing degree of positive
correlation is that twisting cannot be explained only by small
triaxiality (large asphericity) values and a possibly noisy
determination of the shape axis. 

\subsection{Correlation with baryonic disc properties}

The presence of baryons produces rounder dark matter halos than found for the 
corresponding DMO counterpart. 
To better understand this relationship we examine the correlation
between halo shape and baryonic disc properties.  
Looking into the measurements already reported by \cite{auriga} and
\cite{Pakmor17}, we find three baryonic quantities that have the
strongest correlation with DM halo triaxiality: the central gas
density in a sphere of radius $1$ kpc, the disc to total mass ratio, and
the optical radius (the  radius  at 
which  the B-band  surface  brightness  drops  below  25  mag  arcsec$^{−2}$). 

Figure~\ref{fig:disc_correlations} shows the correlations of
these quantities with the triaxiality at five different radii.
We use the Spearmen's rank correlation coefficient to quantify the
correlation strength. 
We find that the strongest correlations are found with the halo shape
measured at radii smaller than $0.12\, R_{200}$.
The  trend is such that halos with large triaxiality correlate with
high gas density, and stellar discs with low mass and small size. 
In turn, massive and large stellar discs with low density of
gas in their cores correlate with low halo triaxiliaty. 

In Figure~\ref{fig:twisting_correlations}, we show the correlation
between twisting and the same baryonic properties that showed a high
degree of correlation with the triaxiality.
In this case the disc to total ratio does not have a significant
degree of correlation. Instead, the disc optical radius
and the gas density at the core are the most correlated. 
The correlation is such that extended stellar discs with a very low
gas density at their cores show the most coherent halo shapes.  

Recently, \citet{2017MNRAS.469..594B} used five zoom simulations 
to report that MW-like DM haloes can be roughly split into two
distinct families: stalled and accreting. 
Stalled haloes assembled early and have remained almost unchanged
in the last $7$ Gyr, while accreting halos have been growing
constantly. \citet{2017MNRAS.469..594B} also link the two populations to
differences in the dark matter filaments feeding the halos. 
Stalled halos at $z=0$ are embedded in a single large thick filament, while
accreting halos are pierced by many thinner filaments. 
To what extent can the twisting and high triaxiality present in the MHD
halos be linked to this dichotomy?

In Figure~\ref{fig:twisting_age} we show the correlation between the
twisting and the age of the bulge, the age of the stellar disc, and the
DM halo age.
We estimate the stellar ages  directly as the mean value in the
stellar age distribution for each component, while the
DM halo age corresponds to the lookback time at which the halo
consolidated half of its present mass.  
In the three cases we find a correlation between age and twisting
that goes in the direction claimed by \citet{2017MNRAS.469..594B}:
components with early assembly (i.e. stalled halos) have a smaller degree of
twisting. 
A detailed study to interpret this with respect to the cosmic 
web location of the halos is however beyond the scope of this paper.

\section{Discussion}
\label{sec:discussion}

\begin{figure*}
\begin{center}
\includegraphics[width=0.6\textwidth]{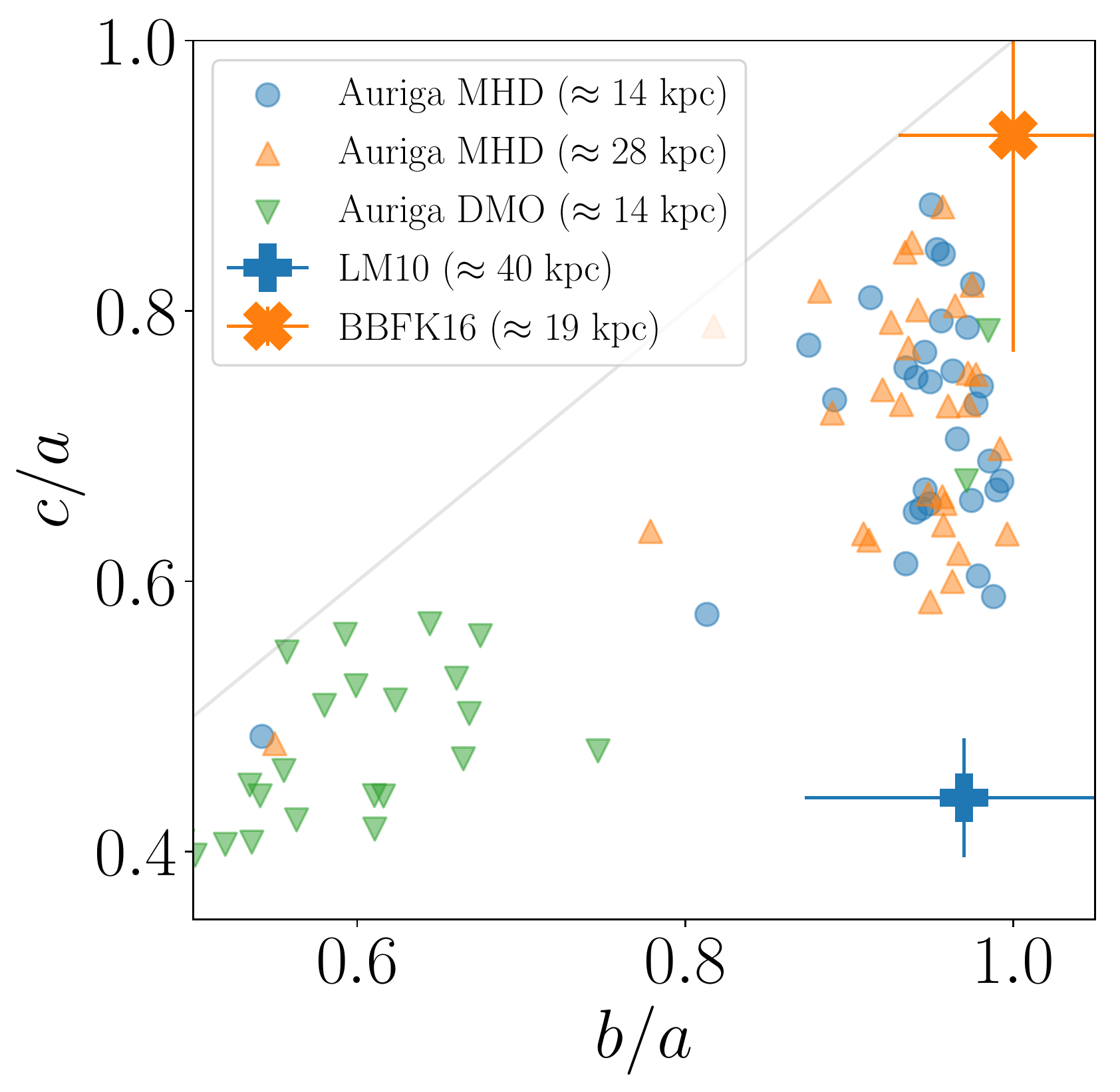}
\end{center}
\caption{Comparison of our results against 
observational constraints for the 
dark matter halo shape in the Milky Way by \citet{LM10} (LM10) and
\citet{Bovy16} (BBFK16).   
We find that 1/5 of the halos in the MHD  sample are consistent with
the constraints by \citet{Bovy16}.
In contrast, none of the halos in the MHD and DMO simulations seems to be
consistent with the results of \citet{LM10}.}
\label{fig:observations}
\end{figure*}

The first prominent effect we found in this paper is that  baryons produce
significantly rounder DM halos. This effect has already been  widely 
reported in the literature, and we confirm it here for the Auriga simulations.
It has also been found that the strength of the change depends on the
numerical resolution, the gas cooling implementation, and the
models describing star formation and stellar feedback
\citep{Bailin05,Debattista08, Bryan13, Butsky16, Chua19, Artale19}.  
The key concept unifying these results is that the baryon distribution
influences and correlates with the dark matter halo shape. 
Indeed, as we demonstrated in the previous section on the correlations
with disc properties, we find a clear tendency that extended
massive stellar discs correlate with spherical dark matter
distributions.  

The second part of our findings concerns the disc-halo alignment.
The dominant trend is that of a strong alignment between
the disc and the dark matter halo shape at all radii. 
In six halos of our sample we find a changing alignment as a function
of radius.
This implies a twisting in the halo shape as a function of
radius.
Considering that this effect is virtually absent in our DMO simulations, it is
very likely that the detailed radial evolution of the alignments
between the disc and the halo depends on the hydrodynamics and
feedback implementations. 

Twisting has been also reported as an uncommon feature in some 
simulated systems found in the literature. 
\cite{JingSuto02} reported it in their dark
matter only simulations as measured through direction changes of large
and medium shape axis. 
They only had three high resolution MW-like halos and could not make a
statistical statement about the significance of this effect.
However, they saw the twisting as an artifact resulting from high
values  of the $b/a$ ratios that make the determination of the medium
axis noisy, an effect that we have demonstrated can be discarded for
our simulations.
\cite{Bailin05} used seven hydrodynamic simulations to study the disc
halo alignment as a function of radius.
For radii smaller than $0.1R_{200}$ they find a strong global
alignment between the halo and the disc.
However, in some halos, at larger radii the angle between the
components twists resulting in a weaker correlation between the disk
and the halo shape. 
A more recent study by \cite{2015MNRAS.453..721V} used the EAGLE simulations to
report the median alignment between the stellar and dark matter
components in $1008$ galaxies around the MW mass range. 
Our results agree on the high degree of alignment on average as a
function of radius.
However, the authors do not mention whether they find any outliers
with the kind of twisting we report here. 

To finalize, we use the results reported by \cite{LM10} and \cite{Bovy16}
to place our results in an observational context.
\cite{LM10} used observations of the Sagittarius tidal stream to
constrain the shape of the gravitational potential.
They point out that previous studies that assumed an
axisymetric galactic potential were not able to fit all the available
dynamic constraints for the Sagittarius stream, motivating 
their approach of using a rigid triaxial potential with coaxial potential
ellipsoids for the dark matter component.  
Their results constrain the triaxiality of this potential
component.  They also translate their results into a triaxiality of the density
contours (which could be compared against our results)
 to be $c/a=0.44$ and $b/a=0.97$ at a radius of $\sim 40$~kpc. 
They do not report any uncertainties for these two values. 
Looking at their plots for the quality of fit criterion as a function
of dark halo axial scales (their Figure 5), we choose a conservative $10\%$
relative uncertainty. One surprising element in their results is that 
the major axis of the
halo shape is perpendicular to the stellar disc plane.  

The results by \cite{Bovy16} are based on the same general approach but use
instead the GD-1 \citep{2006ApJ...641L..37G} and Pal 5 \citep{2009AJ....137.3378O}
streams to constraint the shape of the dark matter component of the
galacic halo potential. They use general models with many degrees of freedom for the galactic
potential in order to measure to what extent these two streams are sensitive
to the triaxiality of the dark matter halo component.
The DM component is written directly as a triaxial density profile
with coaxial ellipsoids, and the corresponding potential is found by
numerical integration. They find that the width of the Pal~5 stream gives $b/a\approx
1$, and therefore fix it to be $b/a=1$ exactly. Using this value they report their 
most stringent constraint of $c/a=0.93\pm0.16$ at a radius of $\approx 19$ kpc from the Galactic
center. 

Figure~\ref{fig:observations} shows an explicit comparison in
the $c/a$--$b/a$ plane of our findings for hydrodynamic simulations against the
results by \cite{LM10} and \cite{Bovy16}.
We find six MHD halos with $b/a<0.93$ and $c/a>0.77$ that could be
considered consistent with the shape constraints  by \citet{Bovy16}, while
only one DMO halo outlier  is consistent with those constraints.
In contrast, none of the simulated halos (MHD nor DMO) is consistent
with the results of \citet{LM10}. The change of triaxiality with radius in 
our simulations cannot account for these two extremely different 
shape constraints at different radii. 

The results by \citet{LM10} would imply that the dark matter halo of our Milky
Way is an extreme outlier in the $\Lambda$CDM model. 
This extreme prolateness also correlates with the extreme triaxiality
of the 11 classical satellites of the MW ($c/a\approx 0.2$, and
$b/a\approx0.9$) measured at larger radii, with a spatial
distribution  also oriented perpendicular to the MW plane, another
highly unusual feature in the $\Lambda$CDM model \citep{2016MNRAS.460.3772S,2018MNRAS.478.5533F,2019MNRAS.488.1166S}. 
An additional caveat is the presence of the Large Magellanic Cloud that might induce a
large enough perturbation (not taken into account in those models) and bias the results
\citep{Vera-Ciro_and_Helmi_2013,2015ApJ...802..128G,2016MNRAS.456L..54P}.

\section{Conclusions}
\label{sec:conclusions}

In this paper we have measured the shape of 30 isolated Milky Way sized
dark matter haloes simulated in the Auriga project using the zoom-in
technique. The halos were simulated using two different setups:
dark matter only (DMO) simulation, and full magnetohydrodynamics (MHD)
including star formation and feedback.
We applied the shape measurement algorithm by \cite{Allgood06} to the
dark matter halos of these simulations to quantify the halo shape as a
function of radius, and the degree of alignment between the angular
momentum in the stellar disc and the dark matter halo shape. 

We find that MHD halos are rounder than DMO halos at all radii.
MHD halos tend towards more oblate shapes, sometimes becoming almost spherical
($T<1/3$), while DMO halos are prone to show more prolate shapes ($T>2/3$).  
The rounding effect by baryons is more noticeable as one moves closer to the galactic
centre, and it strongly  correlates with baryonic properties of the disc.
More precisely, the triaxiality is smaller for extended and massive
stellar discs with low gas densities at their cores.

We also measured the alignment of the halo with the stellar disc's angular
momentum at different radii. For the majority of the sample, the angular momentum
is strongly aligned with the minor axis of the halo at all measured radii.
However, in some halos the alignment changes noticeably with radius. 
This alignment evolution implies a radial twisting between the ellipsoids
describing the halo shape. We quantify this twist with the standard deviation of the angle
between the angular momentum and the halo minor axis at radii below
$\leq 0.5\, R_{200}$, and find that younger bulges and higher gas
densities correlate with larger twisting values. 

We compared our results against two observational constraints for the
dark matter halo shape of the Milky Way. The constraints are at two different radii
coming from different observational tracers. 
We find that $20\%$ of the halos in the MHD simulations are consistent with
the constraints by \cite{Bovy16} at $\approx 19$ kpc,  corresponding to
an almost spherical halo, while none of the halos, neither in MHD nor
DMO, has overlap with the shape constraints by \cite{LM10} at
$\approx 40\,{\rm kpc}$, which argue for a more oblate shape.

A more complete understanding of the influence of baryons on the
different properties we have measured will require at least two more elements.
First, a study on how the halo and the disc co-evolved as a
function of time. 
For instance, using dark matter only simulations, \citet{VeraCiro11} suggested that
the current dark matter halo shape strongly correlates with the time
evolution of the halo as traced by the shape measured at the virial radius. 
This effect might largely be washed out and no longer hold once baryons are included. 
The opposite trend of triaxiality in MHD simulation as a function of
radius compared to the DMO simulations, together with the twisting effect in some of 
the halos, seems to suggest that memory of the historical buildup is at most poorly 
reflected in the shapes at $z=0$.
The second element to take into account is the effect of the cosmic web
(both dark matter and gaseous) on any evolutionary trend
\citep{2014MNRAS.443.1090F, 2017MNRAS.469..594B, 2019MNRAS.487.1607G}.

We also think  that the twisting density shells we find in
some of the halos are a feature that deserves further study, especially
in light of attempts to  constrain halo shape parameters with tidal stream data.  The
inclusion of a parameterization describing this degree of twisting
might relax the present conflicts between the observational inferences and the
numerical results.

\section*{Acknowledgements}
We acknowledge fruitful discussions with Patricia Tissera during the
preparation of the manuscript.
This project has received funding from the European Union's Horizon
2020 Research and Innovation Programme under the Marie
Sk\l{}odowska-Curie grant agreement No 734374. 

\bibliographystyle{mnras}
\bibliography{references}

\begin{thebibliography}{}
\makeatletter
\relax
\def\mn@urlcharsother{\let\do\@makeother \do\$\do\&\do\#\do\^\do\_\do\%\do\~}
\def\mn@doi{\begingroup\mn@urlcharsother \@ifnextchar [ {\mn@doi@}
  {\mn@doi@[]}}
\def\mn@doi@[#1]#2{\def\@tempa{#1}\ifx\@tempa\@empty \href
  {http://dx.doi.org/#2} {doi:#2}\else \href {http://dx.doi.org/#2} {#1}\fi
  \endgroup}
\def\mn@eprint#1#2{\mn@eprint@#1:#2::\@nil}
\def\mn@eprint@arXiv#1{\href {http://arxiv.org/abs/#1} {{\tt arXiv:#1}}}
\def\mn@eprint@dblp#1{\href {http://dblp.uni-trier.de/rec/bibtex/#1.xml}
  {dblp:#1}}
\def\mn@eprint@#1:#2:#3:#4\@nil{\def\@tempa {#1}\def\@tempb {#2}\def\@tempc
  {#3}\ifx \@tempc \@empty \let \@tempc \@tempb \let \@tempb \@tempa \fi \ifx
  \@tempb \@empty \def\@tempb {arXiv}\fi \@ifundefined
  {mn@eprint@\@tempb}{\@tempb:\@tempc}{\expandafter \expandafter \csname
  mn@eprint@\@tempb\endcsname \expandafter{\@tempc}}}

\bibitem[\protect\citeauthoryear{{Abadi}, {Navarro}, {Fardal}, {Babul}  \&
  {Steinmetz}}{{Abadi} et~al.}{2010}]{Abadi10}
{Abadi} M.~G.,  {Navarro} J.~F.,  {Fardal} M.,  {Babul} A.,   {Steinmetz} M.,
  2010, \mn@doi [\mnras] {10.1111/j.1365-2966.2010.16912.x}, \href
  {http://adsabs.harvard.edu/abs/2010MNRAS.407..435A} {407, 435}

\bibitem[\protect\citeauthoryear{{Allgood}, {Flores}, {Primack}, {Kravtsov},
  {Wechsler}, {Faltenbacher}  \& {Bullock}}{{Allgood} et~al.}{2006}]{Allgood06}
{Allgood} B.,  {Flores} R.~A.,  {Primack} J.~R.,  {Kravtsov} A.~V.,  {Wechsler}
  R.~H.,  {Faltenbacher} A.,   {Bullock} J.~S.,  2006, \mn@doi [\mnras]
  {10.1111/j.1365-2966.2006.10094.x}, \href
  {http://adsabs.harvard.edu/abs/2006MNRAS.367.1781A} {367, 1781}

\bibitem[\protect\citeauthoryear{{Artale}, {Pedrosa}, {Tissera}, {Cataldi}  \&
  {Di Cintio}}{{Artale} et~al.}{2019}]{Artale19}
{Artale} M.~C.,  {Pedrosa} S.~E.,  {Tissera} P.~B.,  {Cataldi} P.,   {Di
  Cintio} A.,  2019, \mn@doi [\aap] {10.1051/0004-6361/201834096}, \href
  {https://ui.adsabs.harvard.edu/abs/2019A%26A...622A.197A} {622, A197}

\bibitem[\protect\citeauthoryear{{Bailin} et~al.,}{{Bailin}
  et~al.}{2005}]{Bailin05}
{Bailin} J.,  et~al., 2005, \mn@doi [\apjl] {10.1086/432157}, \href
  {https://ui.adsabs.harvard.edu/abs/2005ApJ...627L..17B} {627, L17}

\bibitem[\protect\citeauthoryear{{Banerjee} \& {Jog}}{{Banerjee} \&
  {Jog}}{2011}]{Banerjee_and_Chanda_2011}
{Banerjee} A.,  {Jog} C.~J.,  2011, \mn@doi [\apjl]
  {10.1088/2041-8205/732/1/L8}, \href
  {http://adsabs.harvard.edu/abs/2011ApJ...732L...8B} {732, L8}

\bibitem[\protect\citeauthoryear{{Borzyszkowski}, {Porciani},
  {Romano-D{\'{\i}}az}  \& {Garaldi}}{{Borzyszkowski}
  et~al.}{2017}]{2017MNRAS.469..594B}
{Borzyszkowski} M.,  {Porciani} C.,  {Romano-D{\'{\i}}az} E.,   {Garaldi} E.,
  2017, \mn@doi [\mnras] {10.1093/mnras/stx873}, \href
  {https://ui.adsabs.harvard.edu/abs/2017MNRAS.469..594B} {469, 594}

\bibitem[\protect\citeauthoryear{{Bovy} \& {Rix}}{{Bovy} \&
  {Rix}}{2013}]{2013ApJ...779..115B}
{Bovy} J.,  {Rix} H.-W.,  2013, \mn@doi [\apj] {10.1088/0004-637X/779/2/115},
  \href {https://ui.adsabs.harvard.edu/abs/2013ApJ...779..115B} {779, 115}

\bibitem[\protect\citeauthoryear{{Bovy}, {Bahmanyar}, {Fritz}  \&
  {Kallivayalil}}{{Bovy} et~al.}{2016}]{Bovy16}
{Bovy} J.,  {Bahmanyar} A.,  {Fritz} T.~K.,   {Kallivayalil} N.,  2016, \mn@doi
  [\apj] {10.3847/1538-4357/833/1/31}, \href
  {http://adsabs.harvard.edu/abs/2016ApJ...833...31B} {833, 31}

\bibitem[\protect\citeauthoryear{{Bowden}, {Evans}  \& {Williams}}{{Bowden}
  et~al.}{2016}]{Bowden_et_al._2016}
{Bowden} A.,  {Evans} N.~W.,   {Williams} A.~A.,  2016, \mn@doi [\mnras]
  {10.1093/mnras/stw994}, \href
  {http://adsabs.harvard.edu/abs/2016MNRAS.460..329B} {460, 329}

\bibitem[\protect\citeauthoryear{{Bryan}, {Kay}, {Duffy}, {Schaye}, {Dalla
  Vecchia}  \& {Booth}}{{Bryan} et~al.}{2013}]{Bryan13}
{Bryan} S.~E.,  {Kay} S.~T.,  {Duffy} A.~R.,  {Schaye} J.,  {Dalla Vecchia} C.,
    {Booth} C.~M.,  2013, \mn@doi [\mnras] {10.1093/mnras/sts587}, \href
  {http://adsabs.harvard.edu/abs/2013MNRAS.429.3316B} {429, 3316}

\bibitem[\protect\citeauthoryear{{Butsky} et~al.,}{{Butsky}
  et~al.}{2016}]{Butsky16}
{Butsky} I.,  et~al., 2016, \mn@doi [\mnras] {10.1093/mnras/stw1688}, \href
  {https://ui.adsabs.harvard.edu/abs/2016MNRAS.462..663B} {462, 663}

\bibitem[\protect\citeauthoryear{{Catena} \& {Ullio}}{{Catena} \&
  {Ullio}}{2010}]{2010JCAP...08..004C}
{Catena} R.,  {Ullio} P.,  2010, \mn@doi [\jcap]
  {10.1088/1475-7516/2010/08/004}, \href
  {https://ui.adsabs.harvard.edu/abs/2010JCAP...08..004C} {8, 004}

\bibitem[\protect\citeauthoryear{{Chua}, {Pillepich}, {Vogelsberger}  \&
  {Hernquist}}{{Chua} et~al.}{2019}]{Chua19}
{Chua} K.~T.~E.,  {Pillepich} A.,  {Vogelsberger} M.,   {Hernquist} L.,  2019,
  \mn@doi [\mnras] {10.1093/mnras/sty3531}, \href
  {https://ui.adsabs.harvard.edu/abs/2019MNRAS.484..476C} {484, 476}

\bibitem[\protect\citeauthoryear{Debattista, Moore, Quinn, Kazantzidis, Maas,
  Mayer, Read  \& Stadel}{Debattista et~al.}{2008}]{Debattista08}
Debattista V.~P.,  Moore B.,  Quinn T.,  Kazantzidis S.,  Maas R.,  Mayer L.,
  Read J.,   Stadel J.,  2008, The Astrophysical Journal, 681, 1076

\bibitem[\protect\citeauthoryear{{Debattista}, {Ro{\v s}kar}, {Valluri},
  {Quinn}, {Moore}  \& {Wadsley}}{{Debattista} et~al.}{2013}]{Debattista13}
{Debattista} V.~P.,  {Ro{\v s}kar} R.,  {Valluri} M.,  {Quinn} T.,  {Moore} B.,
    {Wadsley} J.,  2013, \mn@doi [\mnras] {10.1093/mnras/stt1217}, \href
  {http://adsabs.harvard.edu/abs/2013MNRAS.434.2971D} {434, 2971}

\bibitem[\protect\citeauthoryear{{Deg} \& {Widrow}}{{Deg} \&
  {Widrow}}{2013}]{Deg_and_Widrow_2013}
{Deg} N.,  {Widrow} L.,  2013, \mn@doi [\mnras] {10.1093/mnras/sts089}, \href
  {http://adsabs.harvard.edu/abs/2013MNRAS.428..912D} {428, 912}

\bibitem[\protect\citeauthoryear{{Dubinski}}{{Dubinski}}{1994}]{Dubinski94}
{Dubinski} J.,  1994, \mn@doi [\apj] {10.1086/174512}, \href
  {https://ui.adsabs.harvard.edu/abs/1994ApJ...431..617D} {431, 617}

\bibitem[\protect\citeauthoryear{{Forero-Romero} \& {Arias}}{{Forero-Romero} \&
  {Arias}}{2018}]{2018MNRAS.478.5533F}
{Forero-Romero} J.~E.,  {Arias} V.,  2018, \mn@doi [\mnras]
  {10.1093/mnras/sty1349}, \href
  {https://ui.adsabs.harvard.edu/abs/2018MNRAS.478.5533F} {478, 5533}

\bibitem[\protect\citeauthoryear{{Forero-Romero}, {Contreras}  \&
  {Padilla}}{{Forero-Romero} et~al.}{2014}]{2014MNRAS.443.1090F}
{Forero-Romero} J.~E.,  {Contreras} S.,   {Padilla} N.,  2014, \mn@doi [\mnras]
  {10.1093/mnras/stu1150}, \href
  {https://ui.adsabs.harvard.edu/abs/2014MNRAS.443.1090F} {443, 1090}

\bibitem[\protect\citeauthoryear{{Ganeshaiah Veena}, {Cautun}, {Tempel}, {van
  de Weygaert}  \& {Frenk}}{{Ganeshaiah Veena}
  et~al.}{2019}]{2019MNRAS.487.1607G}
{Ganeshaiah Veena} P.,  {Cautun} M.,  {Tempel} E.,  {van de Weygaert} R.,
  {Frenk} C.~S.,  2019, \mn@doi [\mnras] {10.1093/mnras/stz1343}, \href
  {https://ui.adsabs.harvard.edu/abs/2019MNRAS.487.1607G} {487, 1607}

\bibitem[\protect\citeauthoryear{{Grand} et~al.,}{{Grand}
  et~al.}{2017}]{auriga}
{Grand} R.~J.~J.,  et~al., 2017, \mn@doi [Monthly Notices of the Royal
  Astronomical Society] {10.1093/mnras/stx071}, \href
  {http://adsabs.harvard.edu/abs/2017MNRAS.467..179G} {467, 179}

\bibitem[\protect\citeauthoryear{{Grillmair} \& {Dionatos}}{{Grillmair} \&
  {Dionatos}}{2006}]{2006ApJ...641L..37G}
{Grillmair} C.~J.,  {Dionatos} O.,  2006, \mn@doi [\apjl] {10.1086/503744},
  \href {https://ui.adsabs.harvard.edu/abs/2006ApJ...641L..37G} {641, L37}

\bibitem[\protect\citeauthoryear{{Helmi} \& {White}}{{Helmi} \&
  {White}}{1999}]{1999MNRAS.307..495H}
{Helmi} A.,  {White} S.~D.~M.,  1999, \mn@doi [\mnras]
  {10.1046/j.1365-8711.1999.02616.x}, \href
  {https://ui.adsabs.harvard.edu/abs/1999MNRAS.307..495H} {307, 495}

\bibitem[\protect\citeauthoryear{{Ibata}, {Lewis}, {Martin}, {Bellazzini}  \&
  {Correnti}}{{Ibata} et~al.}{2013}]{Ibata_et_al._2013}
{Ibata} R.,  {Lewis} G.~F.,  {Martin} N.~F.,  {Bellazzini} M.,   {Correnti} M.,
   2013, \mn@doi [\apjl] {10.1088/2041-8205/765/1/L15}, \href
  {http://adsabs.harvard.edu/abs/2013ApJ...765L..15I} {765, L15}

\bibitem[\protect\citeauthoryear{{Iocco}, {Pato}  \& {Bertone}}{{Iocco}
  et~al.}{2015}]{Iocco15}
{Iocco} F.,  {Pato} M.,   {Bertone} G.,  2015, \mn@doi [Nature Physics]
  {10.1038/nphys3237}, \href
  {https://ui.adsabs.harvard.edu/abs/2015NatPh..11..245I} {11, 245}

\bibitem[\protect\citeauthoryear{{Johnston}}{{Johnston}}{1998}]{1998ApJ...495..297J}
{Johnston} K.~V.,  1998, \mn@doi [\apj] {10.1086/305273}, \href
  {https://ui.adsabs.harvard.edu/abs/1998ApJ...495..297J} {495, 297}

\bibitem[\protect\citeauthoryear{{Kazantzidis}, {Abadi}  \&
  {Navarro}}{{Kazantzidis} et~al.}{2010}]{Kazantzidis10}
{Kazantzidis} S.,  {Abadi} M.~G.,   {Navarro} J.~F.,  2010, \mn@doi [\apjl]
  {10.1088/2041-8205/720/1/L62}, \href
  {http://adsabs.harvard.edu/abs/2010ApJ...720L..62K} {720, L62}

\bibitem[\protect\citeauthoryear{{Law} \& {Majewski}}{{Law} \&
  {Majewski}}{2010}]{LM10}
{Law} D.~R.,  {Majewski} S.~R.,  2010, \mn@doi [The Astrophysics Journal]
  {10.1088/0004-637X/714/1/229}, \href
  {http://adsabs.harvard.edu/abs/2010ApJ...714..229L} {714, 229}

\bibitem[\protect\citeauthoryear{{Navarro} \& {Steinmetz}}{{Navarro} \&
  {Steinmetz}}{1997}]{1997ApJ...478...13N}
{Navarro} J.~F.,  {Steinmetz} M.,  1997, \mn@doi [\apj] {10.1086/303763}, \href
  {https://ui.adsabs.harvard.edu/abs/1997ApJ...478...13N} {478, 13}

\bibitem[\protect\citeauthoryear{{Odenkirchen}, {Grebel}, {Kayser}, {Rix}  \&
  {Dehnen}}{{Odenkirchen} et~al.}{2009}]{2009AJ....137.3378O}
{Odenkirchen} M.,  {Grebel} E.~K.,  {Kayser} A.,  {Rix} H.-W.,   {Dehnen} W.,
  2009, \mn@doi [\aj] {10.1088/0004-6256/137/2/3378}, \href
  {https://ui.adsabs.harvard.edu/abs/2009AJ....137.3378O} {137, 3378}

\bibitem[\protect\citeauthoryear{{Olling} \& {Merrifield}}{{Olling} \&
  {Merrifield}}{2000}]{2000MNRAS.311..361O}
{Olling} R.~P.,  {Merrifield} M.~R.,  2000, \mn@doi [\mnras]
  {10.1046/j.1365-8711.2000.03053.x}, \href
  {https://ui.adsabs.harvard.edu/abs/2000MNRAS.311..361O} {311, 361}

\bibitem[\protect\citeauthoryear{{Pakmor} \& {Springel}}{{Pakmor} \&
  {Springel}}{2013}]{2013MNRAS.432..176P}
{Pakmor} R.,  {Springel} V.,  2013, \mn@doi [\mnras] {10.1093/mnras/stt428},
  \href {https://ui.adsabs.harvard.edu/abs/2013MNRAS.432..176P} {432, 176}

\bibitem[\protect\citeauthoryear{{Pakmor} et~al.,}{{Pakmor}
  et~al.}{2017}]{Pakmor17}
{Pakmor} R.,  et~al., 2017, \mn@doi [\mnras] {10.1093/mnras/stx1074}, \href
  {https://ui.adsabs.harvard.edu/abs/2017MNRAS.469.3185P} {469, 3185}

\bibitem[\protect\citeauthoryear{{Pearson}, {K{\"u}pper}, {Johnston}  \&
  {Price-Whelan}}{{Pearson} et~al.}{2015}]{Pearson_et_al._2015}
{Pearson} S.,  {K{\"u}pper} A.~H.~W.,  {Johnston} K.~V.,   {Price-Whelan}
  A.~M.,  2015, \mn@doi [\apj] {10.1088/0004-637X/799/1/28}, \href
  {http://adsabs.harvard.edu/abs/2015ApJ...799...28P} {799, 28}

\bibitem[\protect\citeauthoryear{{Planck Collaboration} et~al.,}{{Planck
  Collaboration} et~al.}{2014}]{2014A&A...571A..16P}
{Planck Collaboration} et~al., 2014, \mn@doi [\aap]
  {10.1051/0004-6361/201321591}, \href
  {https://ui.adsabs.harvard.edu/abs/2014A%26A...571A..16P} {571, A16}

\bibitem[\protect\citeauthoryear{{Schaye} et~al.,}{{Schaye}
  et~al.}{2015}]{Eagle}
{Schaye} J.,  et~al., 2015, \mn@doi [Monthly Notices of the Royal Astronomical
  Society] {10.1093/mnras/stu2058}, \href
  {http://adsabs.harvard.edu/abs/2015MNRAS.446..521S} {446, 521}

\bibitem[\protect\citeauthoryear{{Sofue}, {Honma}  \& {Omodaka}}{{Sofue}
  et~al.}{2009}]{2009PASJ...61..227S}
{Sofue} Y.,  {Honma} M.,   {Omodaka} T.,  2009, \mn@doi [\pasj]
  {10.1093/pasj/61.2.227}, \href
  {https://ui.adsabs.harvard.edu/abs/2009PASJ...61..227S} {61, 227}

\bibitem[\protect\citeauthoryear{{Springel}}{{Springel}}{2010}]{arepo}
{Springel} V.,  2010, \mn@doi [Monthly Notices of the Royal Astronomical
  Society] {10.1111/j.1365-2966.2009.15715.x}, \href
  {http://adsabs.harvard.edu/abs/2010MNRAS.401..791S} {401, 791}

\bibitem[\protect\citeauthoryear{{Tremaine}}{{Tremaine}}{1999}]{1999MNRAS.307..877T}
{Tremaine} S.,  1999, \mn@doi [\mnras] {10.1046/j.1365-8711.1999.02690.x},
  \href {https://ui.adsabs.harvard.edu/abs/1999MNRAS.307..877T} {307, 877}

\bibitem[\protect\citeauthoryear{{Vera-Ciro} \& {Helmi}}{{Vera-Ciro} \&
  {Helmi}}{2013}]{Vera-Ciro_and_Helmi_2013}
{Vera-Ciro} C.,  {Helmi} A.,  2013, \mn@doi [\apjl]
  {10.1088/2041-8205/773/1/L4}, \href
  {http://adsabs.harvard.edu/abs/2013ApJ...773L...4V} {773, L4}

\bibitem[\protect\citeauthoryear{{Vera-Ciro}, {Sales}, {Helmi}, {Frenk},
  {Navarro}, {Springel}, {Vogelsberger}  \& {White}}{{Vera-Ciro}
  et~al.}{2011}]{VeraCiro11}
{Vera-Ciro} C.~A.,  {Sales} L.~V.,  {Helmi} A.,  {Frenk} C.~S.,  {Navarro}
  J.~F.,  {Springel} V.,  {Vogelsberger} M.,   {White} S.~D.~M.,  2011, \mn@doi
  [\mnras] {10.1111/j.1365-2966.2011.19134.x}, \href
  {http://adsabs.harvard.edu/abs/2011MNRAS.416.1377V} {416, 1377}

\makeatother
\end{thebibliography}

\end{document}